\documentclass[11pt]{article} 
\usepackage{mystyle-new}
\usepackage{authblk}
\usepackage[T1]{fontenc}
\usepackage{epsfig,amsmath} 
\usepackage{hepnames,hepunits}
\usepackage{hyperref}
\usepackage{color}
\usepackage{graphicx}
\usepackage{orcidlink}
\definecolor{red}{rgb}{1,0,0}
\def\lesssim{\ \hbox{\raise 2pt \hbox{$<$} \kern -13pt
                     \lower 3pt \hbox{$\sim$}}\ }
\def\greatersim{\ \hbox{\raise 2pt \hbox{$>$} \kern -13pt
                     \lower 3pt \hbox{$\sim$}}\ }

\def\lsim{\mathrel{\rlap{\lower4pt\hbox{\hskip1pt$\sim$}}
    \raise1pt\hbox{$<$}}}                
\def\gsim{\mathrel{\rlap{\lower4pt\hbox{\hskip1pt$\sim$}}
    \raise1pt\hbox{$>$}}}                

\def\cascade{{\sc Cascade}}
\def\pythia{{\sc Pythia}}
\def\herwig{{\sc Herwig}}

\def\mcatnlo{{MCatNLO}}

\input epsf.tex
\def\desepsf(#1 width #2){\epsfxsize=#2 \epsfbox{#1}}
\def\kt{\ensuremath{k_{\rm T}}}

\def\pt{\ensuremath{p_{\rm T}}}
\def\PZ{\ensuremath{Z}}
\def\PZG{\ensuremath{Z/\gamma}}

\def\qt{\ensuremath{q_{\rm T}}}
\def\zdyn{\ensuremath{z_{\rm dyn}}}
\def\ptll{\ensuremath{p_{\rm T}(\ell\ell)}}

\newcommand{\alphas}{\ensuremath{\alpha_\mathrm{s}}}

\newcommand{\mdy}{\ensuremath{m_{\small\text{DY}}}\xspace}

\newcommand{\PBM}{PB}
\newcommand{\PBset}{{PB-NLO-2018}}

\newcommand{\MCatNLO}{{\sc MadGraph5\_aMC@NLO}}
\newcommand{\CAS}{{\mcatnlo+CAS3}}

\newcommand{\GeV}{\text{GeV}\xspace}
\newcommand{\TeV}{\text{TeV}\xspace}

\newenvironment{tolerant}[1]{\par\tolerance=#1\relax}{ \par }
\usepackage{amsmath,bm}
\usepackage{lineno}

\usepackage{cite,./mcite}
\usepackage{tikz}
\usepackage[symbol]{footmisc}

\providecommand{\DOI}[1]{\href{http://dx.doi.org/#1}}

\begin{document}

\title{
The small \boldmath\kt\- region in Drell-Yan production at next-to-leading order with the Parton Branching Method }
\author[1,2]{I.~Bubanja\thanks{itana.bubanja@cern.ch}}
\affil[1]{Faculty of Science and Mathematics, University of Montenegro, Podgorica, Montenegro}
\affil[2]{Interuniversity Institute for High Energies (IIHE), Universit\'e libre de Bruxelles, Belgium} 
\author[3]{A.~Bermudez~Martinez\orcidlink{0000-0001-8822-4727}}
\affil[3]{CERN, Geneva}
\author[2]{L.~Favart\thanks{Laurent.Favart@ulb.be}\orcidlink{0000-0003-1645-7454}}
\author[4]{F.~Guzman\orcidlink{0000-0002-7612-1488}}
\affil[4]{InSTEC, Universidad de La Habana, Havanna, Cuba}
\author[3,5,6]{F.~Hautmann\orcidlink{0000-0001-7563-687X}}
\affil[5]{Elementary Particle Physics, University of Antwerp, Belgium}
\affil[6]{University of Oxford, UK}
\author[7,8]{H.~Jung \thanks{hannes.jung@desy.de}\orcidlink{0000-0002-2964-9845}}
\affil[7]{Deutsches Elektronen-Synchrotron DESY, Germany}\affil[8]{II. Institut f\"ur Theoretische Physik, Universit\"at Hamburg,  Hamburg, Germany}
\author[5]{A.~Lelek\orcidlink{0000-0001-5862-2775}}
\author[7]{M.~Mendizabal\orcidlink{0000-0002-6506-5177}}
\author[7]{K.~Moral~Figueroa\orcidlink{0000-0003-1987-1554}}
\author[9]{L.~Moureaux\thanks{louis.moureaux@cern.ch}\orcidlink{0000-0002-2310-9266}}
\affil[9]{Institut f\"ur Experimentalphysik, Universit\"at Hamburg, Hamburg, Germany}
\author[1]{N.~Raicevic\thanks{natasar@ucg.ac.me}\orcidlink{0000-0002-2386-2290}}
\author[10]{M.~Seidel\orcidlink{0000-0003-3550-6151}}
\affil[10]{Riga Technical University, Riga, Latvia}
\author[7]{S.~Taheri~Monfared\thanks{sara.taheri.monfared@desy.de}\orcidlink{0000-0003-2988-7859}}

\date{}
\begin{titlepage} 
\maketitle
\vspace*{-18cm}
\begin{flushright}
DESY-23-209\\
\today
\end{flushright}
\end{titlepage}

\begin{abstract}
The Parton Branching (\PBM ) method describes the evolution of transverse momentum dependent (TMD) parton distributions,  covering all kinematic regions from small to large transverse 
momenta \kt . The small \kt -region is very sensitive both to the contribution of the intrinsic  motion of partons (intrinsic \kt ) and to the resummation of  soft gluons taken into account by the PB TMD evolution equations.  
We study the role of soft-gluon emissions in TMD as well as integrated parton distributions.

We perform a detailed investigation of the 
\PBM\ TMD methodology at next-to-leading order (NLO) in Drell-Yan (DY) production for low transverse momenta. We present the extraction of  the  nonperturbative 
``intrinsic-\kt\ '' distribution from recent measurements of  DY transverse momentum distributions at the LHC across a wide range in DY masses,  including a detailed treatment of statistical, correlated and uncorrelated uncertainties.  We comment on the (in)dependence of  intrinsic transverse momentum on DY mass and center-of-mass energy, and on the comparison with other approaches.

\end{abstract} 

\section{Introduction} \label{Intro}
The measurement of the vector boson transverse momentum, \pt ,
in Drell-Yan (DY) production~\cite{Drell:1970wh} allows one to
investigate in detail many different aspects of the strong interaction sector of the 
Standard Model, and their impact on precision electroweak measurements.
The very low $p_T$ region of the DY cross section is 
sensitive to the contribution from the nonperturbative transverse motion of partons inside the hadrons; 
additionally at low transverse momentum multiple 
soft gluon emissions have to be resummed; at larger 
transverse momenta perturbative higher-order 
contributions become dominant. 
The precise description of the \PZG\ boson transverse momentum distribution has been investigated 
since the 1980's, and approaches like CSS~\cite{Collins:1984kg} analytic resummation 
and parton-shower
~\cite{Sjostrand:2014zea,Bellm:2015jjp,Bahr:2008pv,Gleisberg:2008ta} numerical algorithms have been applied with different success. 

 In this work we explore the approach~\cite{Martinez:2019mwt,Martinez:2020fzs}  
to DY \pt\ spectra based on the parton branching (PB) TMD methodology in momentum space 
proposed in~\cite{Hautmann:2017fcj,Hautmann:2017xtx}, and we perform a detailed analysis of the 
small-\pt\ region for wide ranges in center of mass energies and in DY masses.
Though fitted only on deep-inelastic scattering (DIS) data 
from HERA experiments, the \PBM -TMD methodology
has been shown to be capable of describing DY \pt\ spectra 
at LHC energies~\cite{Martinez:2019mwt} 
and at low energies~\cite{Martinez:2020fzs} 
without any need for adjustment of parameters. 
This approach takes into account simultaneously soft gluon radiations 
and the transverse momentum recoils in the parton branchings along the QCD cascade. It provides a successful 
natural treatment of the multiple-scale problem of the  DY transverse momentum for transverse momenta much smaller than DY masses but 
also of the DY with hard jet production~\cite{BermudezMartinez:2021lxz}. It also confirms the universality of the TMDs 
being able to describe both DIS and DY cross sections at all available center of mass energies \cite{Angeles-Martinez:2015sea}.
Alternative approaches based on parton showers in standard Monte Carlo event generators 
like \pythia8 \cite{Sjostrand:2014zea} can also describe multi-differential DY cross section but it has been observed
that they require intrinsic transverse momentum distributions strongly dependent 
on $\sqrt{s}$ ~\cite{Gieseke:2007ad,Sjostrand:2004pf}. 
In order to describe the measurements at  LHC energies, a Gaussian width exceeding the Fermi motion kinematics is needed. 
Approaches based on CSS~\cite{Isaacson:2023iui} provide very precise analytic predictions for inclusive enough
observables like the Drell-Yan cross section transverse momentum.
 In this paper we study in detail the low \kt\ behavior of the \PBM -TMD parton distributions where both 
very soft gluon emission and intrinsic-\kt\ contribute significantly and interplay.
The results presented here provide a
multi-scale economical and coherent approach demonstrating the sensitivity to nonperturbative TMD contributions and 
first steps in disentangling the intrinsic-\kt\ contribution from the nonperturbative Sudakov one  \cite{Hautmann:2020cyp}.
We compare DY theoretical predictions with experimental measurements in wide ranges in center-of-mass energies, 
$\sqrt{s}$ and in DY masses, \mdy , to extract the intrinsic-\kt\ parameter from the transverse momentum distributions.
We are carefully taking into account systematic and statistical uncertainties using the 
breakdown of experimental uncertainties provided by the full set of covariance matrices available in the recent Drell-Yan
differential cross section measurement at 13 TeV \cite{CMS:2022ubq} 
and we treat for the first time the scale uncertainties in the theoretical predictions as correlated uncertainties
within a given mass bin.

The results for TMD parameters such as intrinsic-\kt\ obtained from the DY analysis in this paper can be compared with
analogous results obtained from TMD fits in the CSS coordinate-space framework,
see e.g.~the recent studies~\cite{Bacchetta:2022awv,Bury:2022czx}.
A significant difference between these approaches and the approach of this
paper concerns the treatment of collinear parton distribution functions (PDFs).
As shown in Refs.~\cite{Martinez:2019mwt,Martinez:2020fzs}, in the approach of this paper the inclusive DGLAP limit is
recovered and fits of collinear distributions are made, e.g.~from inclusive DIS
structure functions, along with TMD distributions~\cite{Martinez:2018jxt,Jung:2021mox,Jung:2021vym}. 
In contrast, CSS approaches do not recover inclusive DGLAP and rather
use an ansatz based on the operator product expansion of TMD
distributions in terms of collinear PDFs, assuming  collinear PDFs to be given
by standard PDF sets. The PDF bias effect~\cite{Bury:2022czx} which results
from this has been shown to influence significantly the central values of the
extracted distributions and dominate the systematic uncertainties in all the
existing TMD determinations based on CSS approaches. The possibility to
treat collinear and TMD distributions on the same footing and determine them
without having to rely on existing PDF fits is a distinctive feature 
of the \PBM\ TMD approach. We believe that in the long run this could
bring significant advantages in pursuing TMD phenomenology.

On the other hand, the results of this paper for intrinsic-\kt\ can also be compared with the case of
parton shower Monte Carlo event generators, such as {\sc Pythia}~\cite{Sjostrand:2014zea}
and {\sc Herwig}~\cite{Bellm:2015jjp}.
Monte Carlo tuning to experimental data shows that parton shower
approaches require intrinsic-\kt\ distributions dependent on the center-of-mass
energy $\sqrt{s}$~\cite{Sjostrand:2004pf,Gieseke:2007ad}, and
a Gaussian width exceeding the Fermi motion kinematics.
In contrast, in the approach of this paper we find that the width of the intrinsic-\kt\ distribution 
has a much milder center-of-mass energy $\sqrt{s}$ dependence.
We obtain more natural Gaussian width $\sigma$,  
$ \sigma  =  q_s / \sqrt{2} $, with $q_s $ close to 1 GeV resulting
from fits to DY measurements from fixed-target to LHC energies.
We propose in this paper that the different behavior, concerning intrinsic-\kt\ distributions, between \PBM\ TMD and
parton-shower approaches  can be ascribed to the different treatment of the contributions to parton evolution from the
nonperturbative Sudakov region, near the soft-gluon resolution boundary.
See also~\cite{BermudezMartinez:2020tys} for a discussion of this and comparison of \PBM\ TMD and parton-shower results.

The paper is organized as follows. In Sec.~\ref{PBTMD} we briefly recall the basic elements of the calculational
framework~\cite{Hautmann:2017fcj,Hautmann:2017xtx,Martinez:2019mwt,BermudezMartinez:2020tys,Jung:2021mox,Jung:2021vym}: 
we start with the \PBM\ TMD approach; next we give a few comments on the treatment of the small transverse momentum
region in this approach; then  we discuss the Monte Carlo computation of DY differential distributions. 
Sec.~\ref{sec:pt} is the central section of the paper, in which we perform fits to DY data and present results for
the intrinsic-\kt\ TMD parameter. We give conclusions in Sec.~\ref{sec:concl}.

\section{PB TMDs and DY production} \label{PBTMD}

To study the different contributions to the low-\pt\ spectrum, at different \mdy\ and different $\sqrt{s}$, 
we calculate DY production cross section in the \PBM\ TMD method, 
which proceeds as described in Refs.~\cite{BermudezMartinez:2020tys,Martinez:2019mwt}.
NLO hard-scattering matrix elements are obtained from the {\scshape MadGraph5\_aMC@NLO}~\cite{Alwall:2014hca}
at next-to-leading (NLO) event generator and matched with TMD parton distributions and showers 
obtained from \PBM\ evolution~\cite{Martinez:2018jxt,Hautmann:2017fcj,Hautmann:2017xtx},
 using the
subtractive matching procedure proposed in~\cite{Martinez:2019mwt}  
and further analyzed in~\cite{Yang:2022qgk}.

We will show that the application of \PBM\ TMD distributions
leads to a non negligible contribution of pure intrinsic-\kt , even if most of the small-\kt\ contribution comes from
the \PBM-evolution. We also show that the proper treatment of photon radiation from the DY decay leptons is rather important, 
especially in the DY mass region below the \PZ\ boson peak. The contribution of intrinsic-\kt\ of heavy flavor partons is found to
be negligible over the whole range since heavy quarks are not present in the initial configuration of the proton.


\subsection{TMD distributions from the PB method} \label{pbtmd}

The PB evolution equations for TMD parton distributions
$ {\cal A}_a ( x , {\bf k } , \mu^2) $
of flavor $a$ are given by~\cite{Hautmann:2017fcj}
\begin{eqnarray}
\label{evoleqforA}
   { {\cal A}}_a(x,{\bf k}, \mu^2)
 &=&
 \Delta_a (  \mu^2  ) \
 { {\cal A}}_a(x,{\bf k},\mu^2_0)
 + \sum_b
 \int {{d^2 {\bf q}^{\prime } } \over {\pi {\bf q}^{\prime 2} } }
 \
{
{\Delta_a (  \mu^2  )}
 \over
{\Delta_a (  {\bf q}^{\prime 2}
 ) }
}
\ \Theta(\mu^2-{\bf q}^{\prime 2}) \
\Theta({\bf q}^{\prime 2} - \mu^2_0)
 \nonumber\\
&\times&
\int_x^{z_M} {{dz}\over z} \;
P_{ab}^{(R)} (\alpha_s
,z)
\;{ {\cal A}}_b\left({x \over z}, {\bf k}+(1-z) {\bf q}^\prime ,
{\bf q}^{\prime 2}\right)
  \;\;  ,
\end{eqnarray}
where $\bf k$ and $\bf q$ are 2-dimensional momentum vectors,  $z_M$ is the soft resolution scale~\cite{Hautmann:2017xtx}, $z$~is the longitudinal momentum
transferred at the branching, $P_{ab}^{(R)} (\alpha_s , z ) $ are the
resolvable splitting functions\footnote{Using transverse momentum dependent splitting functions as described in Ref.\cite{Hautmann:2022xuc} would require using off-shell matrix elements and a completely new fit to inclusive structure functions.}  (whose explicit expressions for all flavor channels are given in~\cite{Hautmann:2017fcj}), and
$ \Delta_a$ are the Sudakov form factors
\begin{equation}
\label{sud-def}
 \Delta_a ( z_M, \mu^2 , \mu^2_0 ) =
\exp \left(  -  \sum_b
\int^{\mu^2}_{\mu^2_0}
{{d {\bf q}^{\prime 2} }
\over {\bf q}^{\prime 2} }
 \int_0^{z_M} dz \  z
\ P_{ab}^{(R)}\left(\alpha_s ,
 z \right)
\right)
  \;\; .
\end{equation}
The branching evolution (\ref{evoleqforA})
fulfills soft-gluon angular
ordering~\cite{Webber:1986mc,Marchesini:1987cf,Catani:1990rr}, with the branching variable
${\bf q}^{\prime 2}$ being related to the transverse
momentum \qt\ of the parton emitted at the branching by
\begin{equation}
\label{angord}
 \qt =  (1-z) \, | {\bf q}^{\prime } | .
\end{equation}
It is shown in~\cite{Hautmann:2017xtx} that angular ordering is essential for the TMD distribution 
arising from the solution of Eq.~(\ref{evoleqforA})  to be  well-defined and independent of the
choice of the soft-gluon resolution scale $z_M = 1 - \varepsilon$ for
$\varepsilon \to 0$. In contrast, $p_T$ ordering  leads, for instance,
to ambiguities in the definition of the
TMD from the $z \to 1$ region.

Analogously to the case of ordinary (collinear) parton distribution functions,
the distribution $  { {\cal A}}_a(x,{\bf k},\mu^2_0)   $
at the starting scale $\mu_0$ of the evolution, in the first term on the right hand side of
Eq.~(\ref{evoleqforA}),  is a nonperturbative boundary condition to the evolution equation, and is
to be determined from experimental data. For simplicity
we parameterize  $ { {\cal A}}_a(x,{\bf k},\mu^2_0) $ in the form
\begin{equation}
\label{TMD_A0}
{\cal A}_{0,a} (x, {\bf k},\mu_0^2)   =  f_{0,a} (x,\mu_0^2)
\cdot \exp\left(-| {\bf k}|^2 / 2 \sigma^2\right) / ( 2 \pi \sigma^2) \; ,
\end{equation}
with the width of the Gaussian distribution given by $ \sigma  =  q_s / \sqrt{2} $,
independent of parton flavor  and $x$, where $q_s$ is the intrinsic-\kt\ parameter.

The scale at which the strong coupling $\alpha_s$ is to be evaluated in
Eqs.~(\ref{evoleqforA}) and (\ref{sud-def}) is a function of the branching variables.
Two scenarios are studied in Refs.~\cite{Martinez:2018jxt,Hautmann:2017fcj}: 
\begin{eqnarray} 
 \label{zlimit}
 && {\rm i})  : \;\;\;   \alpha_s = \alpha_s ({\bf q}^{\prime 2})
 \nonumber\\
 && {\rm ii}) :  \;\;\;  \alpha_s = \alpha_s ({\bf q}^{\prime 2} (1-z)^2) = \alpha_s (\qt^2)
\end{eqnarray}
In scenario i), it is shown in \cite{Hautmann:2017fcj} that Eq.~(\ref{evoleqforA}), in the collinear case, i.e.\ once it is integrated 
over all transverse momenta, reproduces exactly the DGLAP 
evolution \cite{Gribov:1972ri,Lipatov:1974qm,Altarelli:1977zs,Dokshitzer:1977sg} of parton densities.
In scenario ii), it is discussed in~\cite{Hautmann:2019biw}
how, upon integration over transverse momenta and suitable treatment of the resolution scale,
Eq.~(\ref{evoleqforA}) returns the CMW coherent branching evolution~\cite{Catani:1990rr}.

In Ref.~\cite{Martinez:2018jxt}, 
fits to precision DIS HERA measurements~\cite{Abramowicz:2015mha}
based on Eqs.~(\ref{evoleqforA}) and (\ref{TMD_A0}), combined with NLO DIS matrix elements, are
performed for both scenarios i)  and  ii),
using the fitting platform \verb+xFitter+~\cite{xFitterDevelopersTeam:2022koz,Alekhin:2014irh}.
It is found that fits to DIS measurements with good $\chi^2$
values can be achieved in either case. Correspondingly, PB-NLO-HERAI+II-2018~set 1 (abbreviated as \PBset~Set1)
(with  the DGLAP-type $ \alpha_s ({\bf q}^{\prime 2}) $)
and PB-NLO-HERAI+II-2018~set2 (abbreviated as \PBset~Set2) (with the angular-ordered CMW-type
$ \alpha_s  (\qt^2)$) are obtained, both
having intrinsic-\kt\ parameter in Eq.~(\ref{TMD_A0}) set to $q_s = $ 0.5 GeV~\cite{Martinez:2018jxt}. All \PBM\ TMD parton distributions (and many others) are accessible in TMDlib and via the graphical 
interface TMDplotter~\cite{Hautmann:2014kza,Abdulov:2021ivr}.

 On the other hand, it is found that \PBset~Set2 provides a much better description, compared to
\PBset~Set1, of measured $Z/\gamma$ transverse momentum spectra 
at the LHC~\cite{Martinez:2019mwt}, 
in low-energy experiments~\cite{BermudezMartinez:2020tys}, and
of di-jet azimuthal correlations near the back-to-back region at the LHC~\cite{Abdulhamid:2021xtt}.
This underlines the relevance of the angular-ordered coupling
$ \alpha_s (\qt^2)$ in regions dominated by soft-gluon emissions.

Based on this observation, in the following we will focus on the \PBset~Set2 approach 
and perform fits to DY  transverse momentum measurements 
to investigate the sensitivity of these measurements to
the nonperturbative TMD intrinsic-\kt\ parameter $q_s$, and
perform determinations of its value.

As discussed in ~\cite{Martinez:2018jxt,Martinez:2019mwt}, 
in order to
complete the definition of the \PBset~Set2 scenario the treatment of the coupling
$\alpha_s$ needs to be specified in the region of small transverse
momenta $\qt \lesssim  q_0$, where $q_0 $ is a semi-hard scale on the order of a GeV.
As in~\cite{Martinez:2018jxt,Martinez:2019mwt}, 
we take
\begin{equation}
\label{freeze}
 \alpha_s = \alpha_s(\max(q^2_{0},\qt^2)),
\end{equation}
setting $q_{0} = $ 1 GeV, which may be regarded as similar in spirit to the ``pre-confinement'' 
proposal in the context of infrared-sensitive QCD processes~\cite{Amati:1980ch,Bassetto:1983mvz} \footnote{Different forms of the extension to small \qt\ could be considered. However, this will entail new fits both to precision DIS data and to DY data.}. 
In the present study, we will perform a determination
of the nonperturbative TMD parameter $q_s$ from DY transverse
spectra by assuming the above behavior for $\alpha_s$. 

To better illustrate the underlying physical picture, we give next a few further comments on nonperturbative
contributions and the treatment of the small transverse momentum region in the PB TMD approach.


As implied by Eqs.~(\ref{evoleqforA}) and (\ref{sud-def}), the PB TMD method incorporates Sudakov evolution
via phase space integrations of appropriate kernels over the resolvable region, i.e.\ over
momentum transfers $z$ up to the soft-gluon resolution scale $z_M$. For each branching
evolution scale ${\bf q}^{\prime 2}$, it is instructive to examine separately
parton emissions  with transverse momenta above the
semi-hard scale $q_0$, $\qt > q_0$, and below $q_0$, $\Lambda_{\rm{QCD}} < \qt \lsim q_0$.
Using the angular ordering relation (\ref{angord}), these emissions are mapped
respectively on the regions
\begin{eqnarray}
 \label{regions-ab}
 && ({a})  : \;\;\;   z < z_{\rm{dyn}} = 1 - q_0 / | {\bf q}^{\prime } |  ,
 \nonumber\\
 && ({b}) :  \;\;\;  z_{\rm{dyn}}  \lsim z < z_M ,
\end{eqnarray}
where $ z_{\rm{dyn}} = 1 - q_0 / | {\bf q}^{\prime } |$ is the dynamical resolution scale
associated with the angular
ordering~\cite{Webber:1986mc,Marchesini:1987cf,Catani:1990rr,Hautmann:2019biw}.
In region $(a)$, the strong coupling (\ref{freeze}) is evaluated at
the scale of the emitted transverse momentum, $ \alpha_s  (\qt^2)$;
the contribution from region $(a)$ to the evolution in Eqs.~(\ref{evoleqforA}), (\ref{sud-def})
corresponds to the perturbative Sudakov resummation (see e.g.~\cite{vanKampen:2021oxe,PB-NNLL}).
In region $(b)$, the strong coupling (\ref{freeze}) freezes around the semi-hard scale $q_0$;  the
contribution from region $(b)$ to the evolution is the  nonperturbative Sudakov form factor
in the PB TMD approach.

It is worth noting that the  \PBset~Set 2 framework provides a very natural and economical description of nonperturbative
Sudakov effects, based on perturbative modeling of the Sudakov form factor (\ref{sud-def})
combined with the infrared $\alpha_s $ behavior (\ref{freeze}):
it does not contain any additional nonperturbative functions and parameters, besides the scale $q_0$.

\begin{figure}[h!tb]
\begin{center} 
\vskip -3cm
\includegraphics[clip,width=0.51\textwidth]{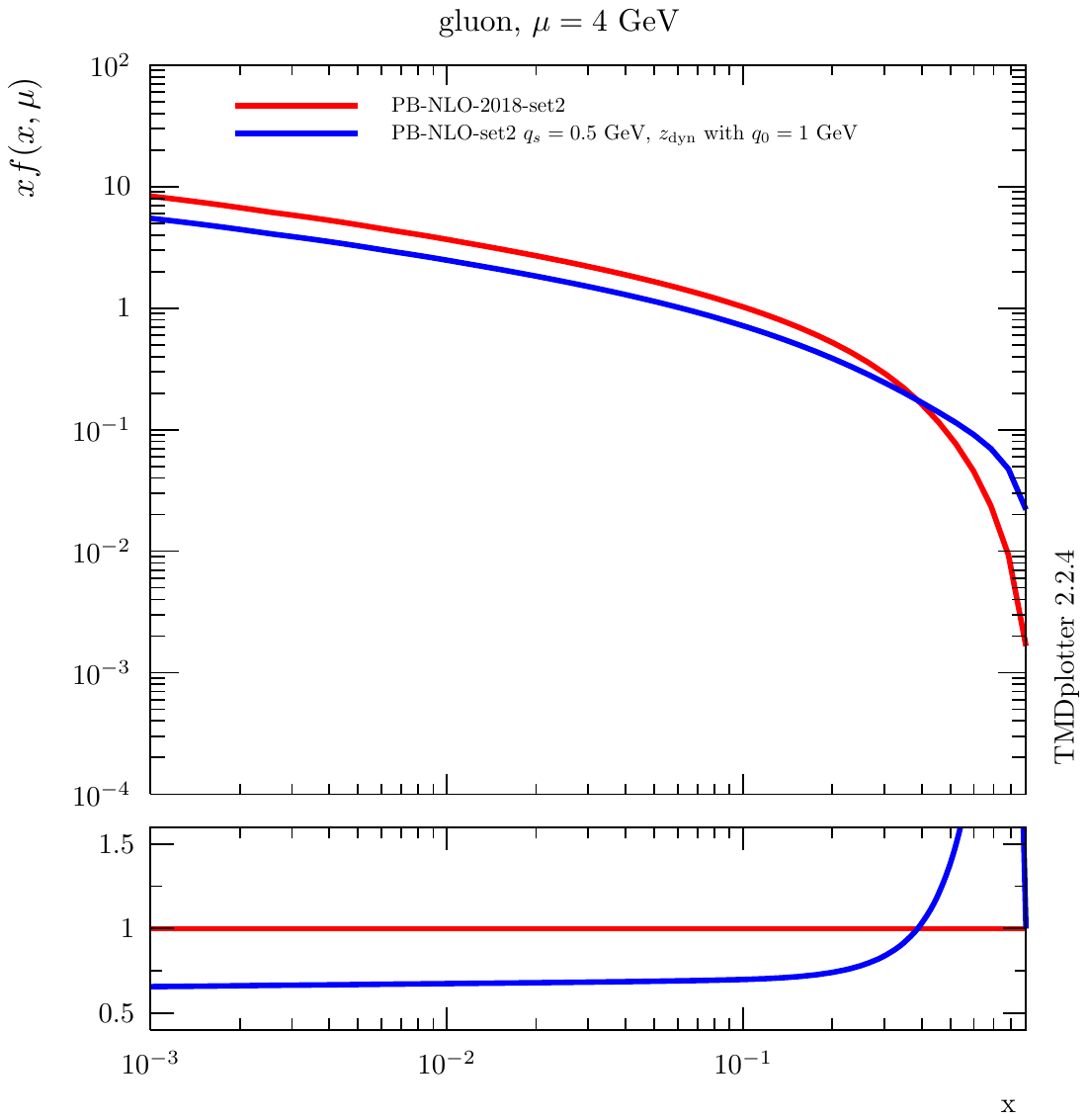} \hskip -0.8cm
\includegraphics[clip,width=0.51\textwidth]{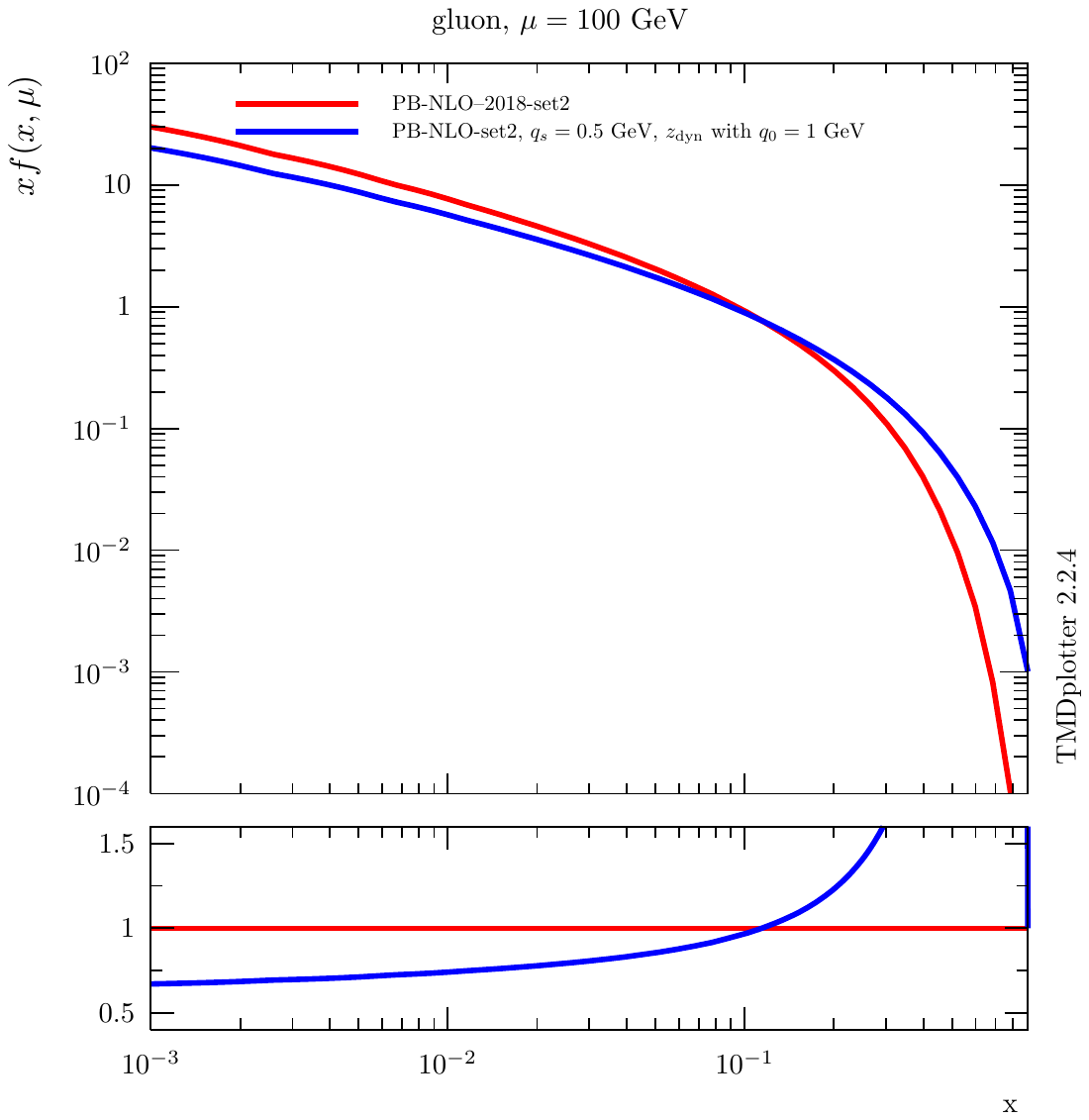}
\vskip -3.5cm
\includegraphics[clip,width=0.51\textwidth]{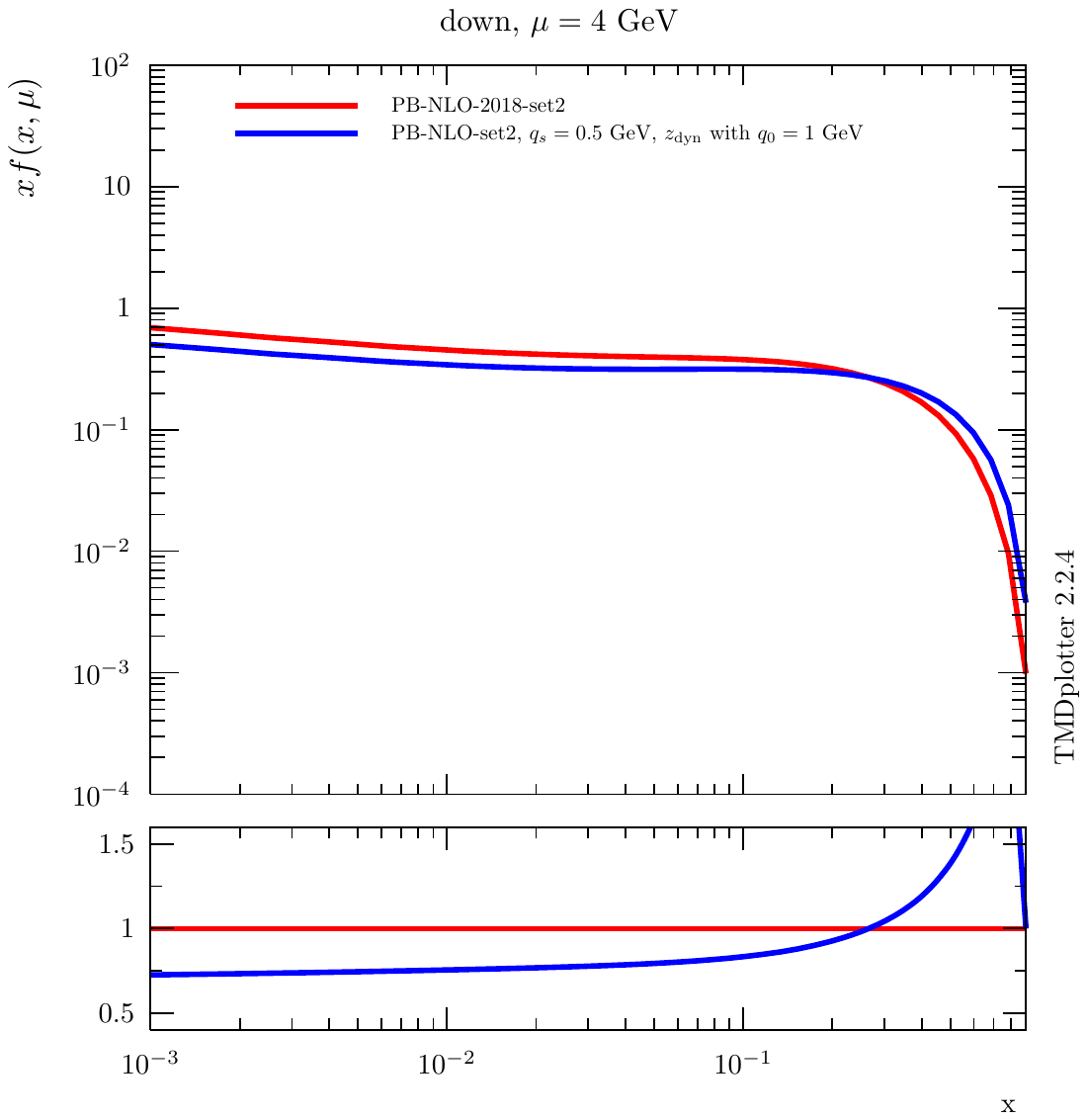} \hskip -0.8cm
\includegraphics[clip,width=0.51\textwidth]{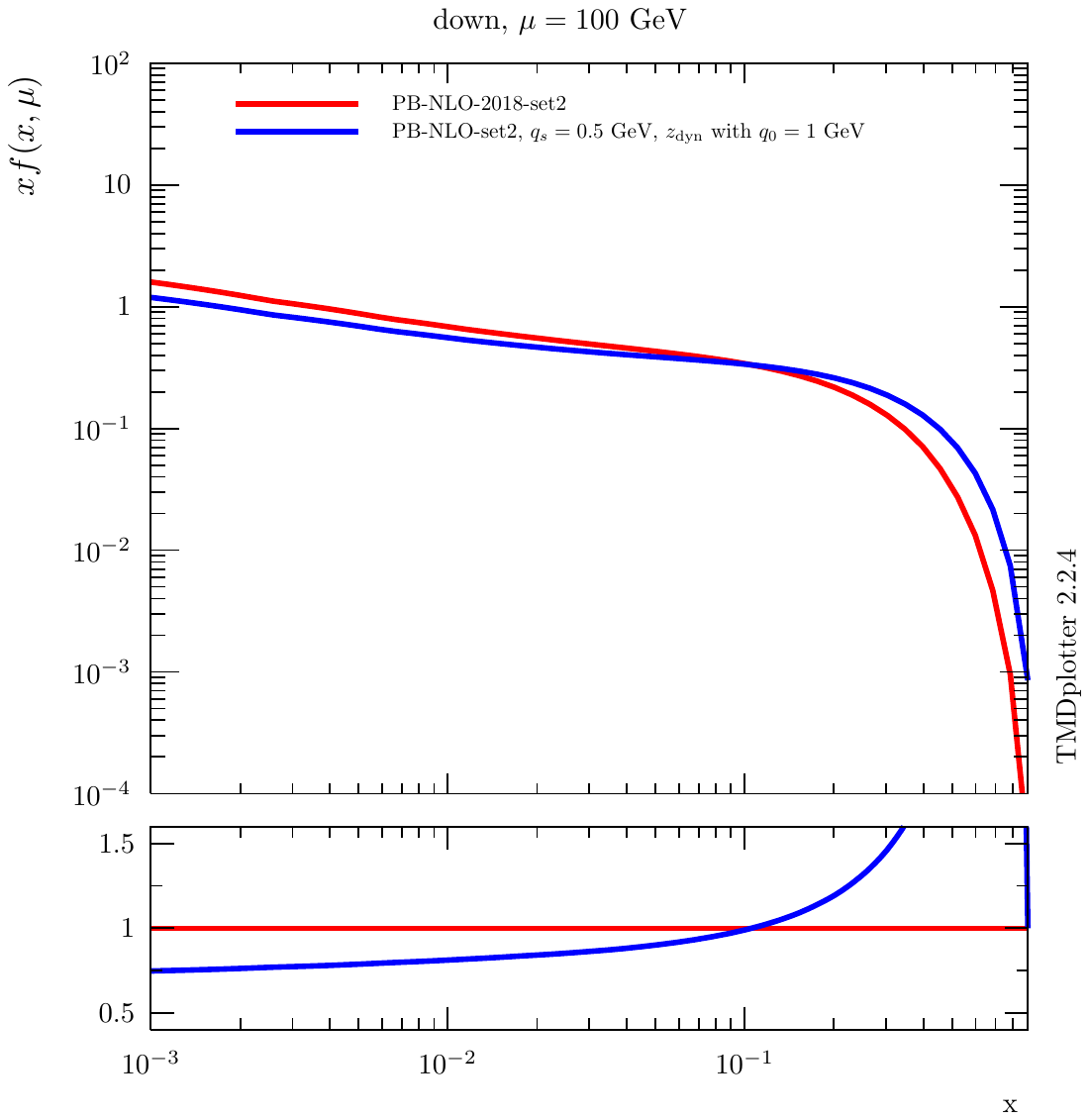}
\vskip -1cm
\caption{\small Integrated gluon and down-quark distributions at $\mu = 4$~GeV (left column) and $\mu = 100$~GeV (right column) 
obtained from the \protect\PBM\ approach based on \PBset~Set2. The red curve is the published \PBset~Set2 \protect\cite{Martinez:2018jxt} and corresponds  to $z_M \to 1$ (regions $(a+b)$ in text)..
The blue curve corresponds to $z_M=\zdyn$ with $q_0=1$~GeV (region $(a)$ only). 
The ratio plots show the ratios to the one for $z_M \to 1$. }
\label{PB-DGLAP}
\end{center}
\end{figure} 

In Fig.~\ref{PB-DGLAP} we show parton distributions obtained with the \PBM\ approach using the starting distributions from  
\PBset~Set2.  
We show distributions for the gluon and down quark parton densities for different values of $z_M$: $z_M \to 1$
(default - regions $(a+b)$ - red curve  \cite{Martinez:2018jxt}) and $z_M=\zdyn = 1 - q_0/{\bf q}^{\prime }$ (region $(a)$ only - blue curve obtained with the 
same parameters as \PBset~Set2 except $z_M$ using {\sc updfevolv } \cite{PBevolution}).
The distributions obtained from \PBset~set2 with $z_M \to 1 $ are significantly different from those applying $z_M=\zdyn $, illustrating the importance of soft contributions even for collinear distributions.
In  Ref.~\cite{Nagy:2020gjv} it was found that limiting the $z$-integration leads to inconsistencies. In Ref.~\cite{Frixione:2023ssx} a procedure to correct 
the $z$ limitation is discussed. A detailed discussion on the role of soft gluons and the nonperturbative Sudakov form factor is given in Ref.~\cite{Mendizabal:2023mel}. 
Please note that the intrinsic-\kt\ distribution, since not part of the collinear calculation, does not affect the collinear parton densities.

\begin{figure}[h!tb]
\begin{center} 
\vskip -2cm
\includegraphics[width=0.51\textwidth]{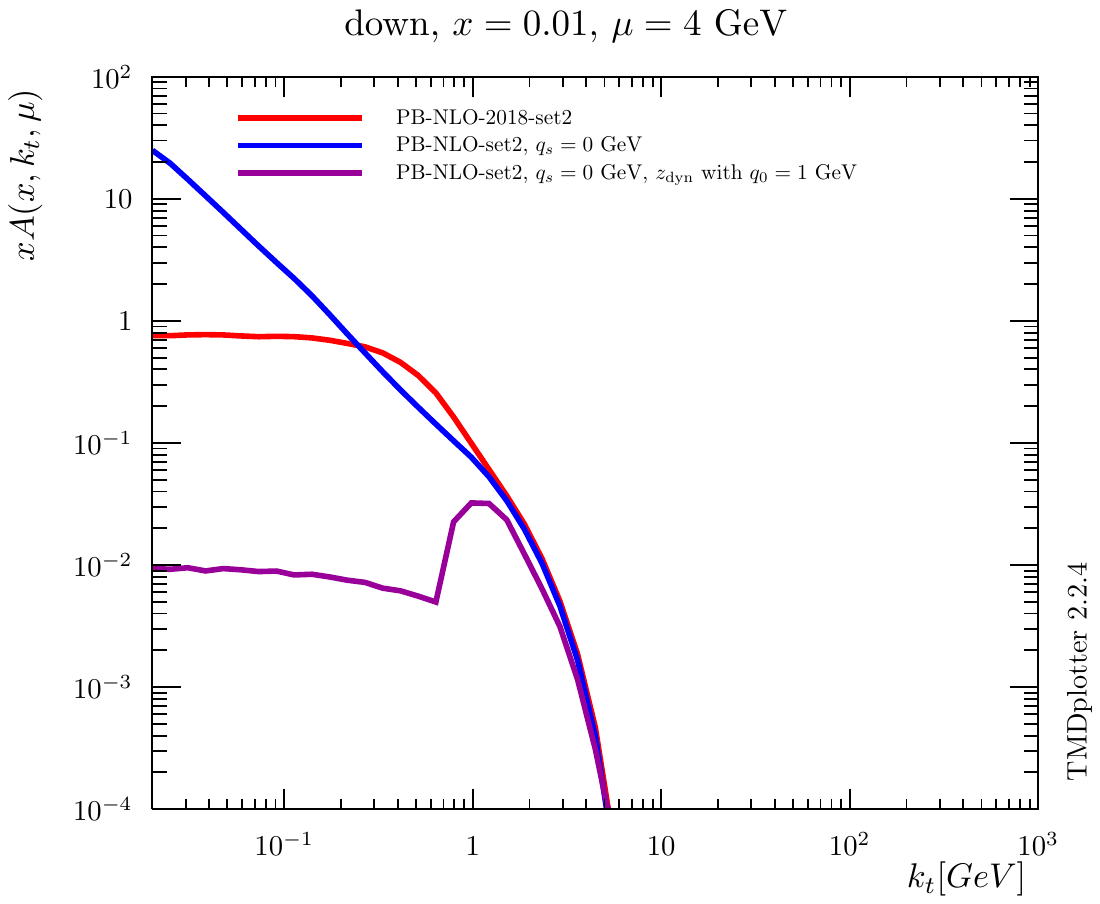} \hskip -0.8cm
\includegraphics[width=0.51\textwidth]{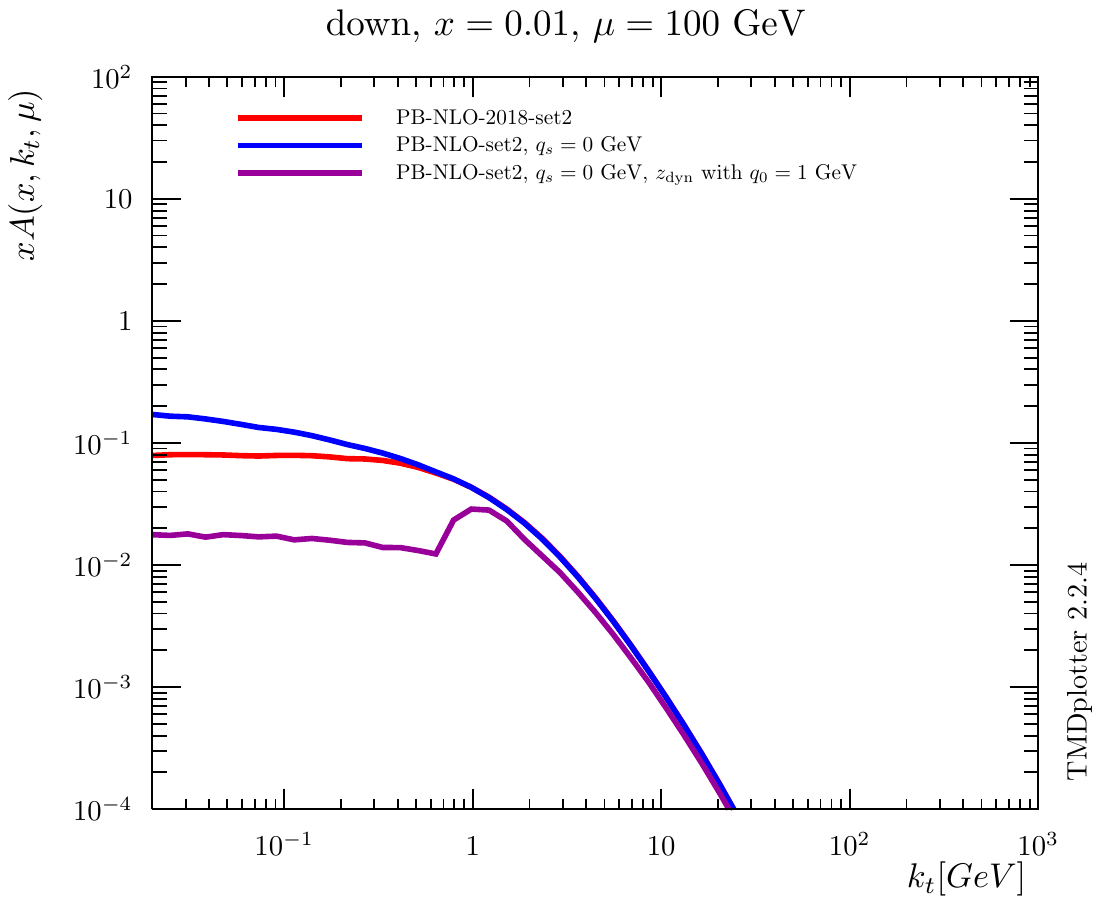}
\vskip -4.5cm
\includegraphics[width=0.51\textwidth]{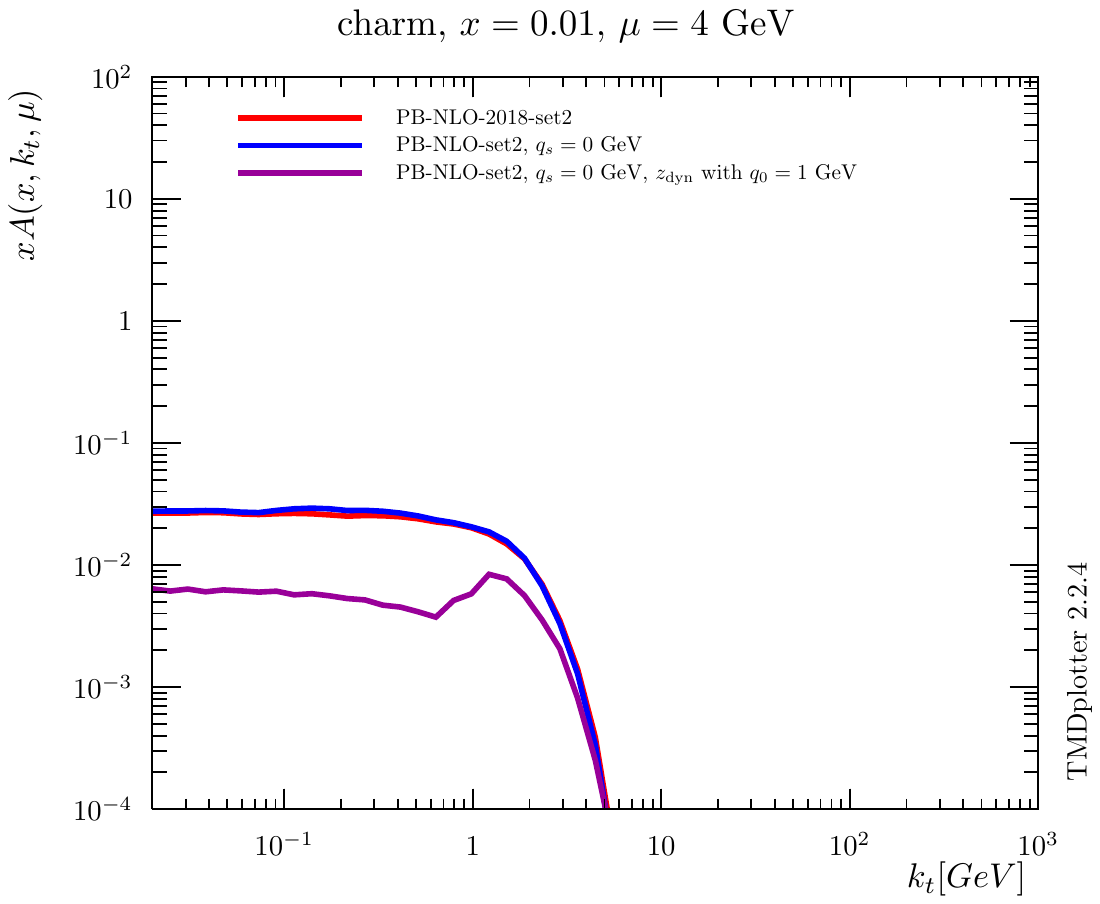} \hskip -0.8cm
\includegraphics[width=0.51\textwidth]{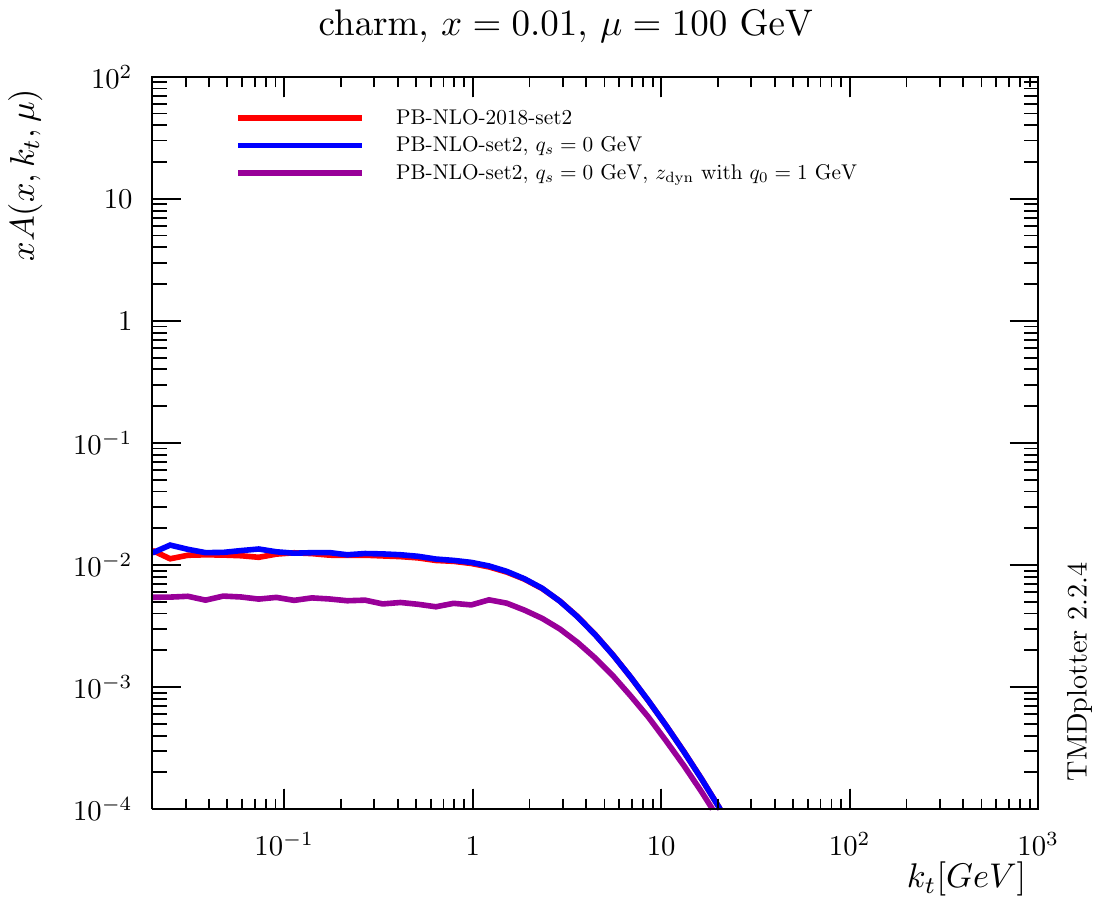}
\vskip -2cm
\caption{\small Transverse momentum distributions of down and charm quarks at $\mu  = 4$~GeV (left column) and $\mu = 100$ GeV 
(right column) obtained from the \protect\PBM\ approach based on \protect\PBset~Set2. 
Two distributions do not include intrinsic-\kt\ : the blue curve corresponds to $z_M \to 1$ (regions a+b in text) 
and the magenta curve to $z_M=\zdyn = 1 - q_0/q$ (region a only). 
The red curve is the published one \PBset~Set2 \protect\cite{Martinez:2018jxt} and including intrinsic-\kt\ and $z_M \to 1$. 
}
\label{TMD-DGLAP}
\end{center}
\end{figure}

In the transverse momentum distributions obtained with the \PBM -approach, the effect of the $z_M$ cut-off is even more visible. 
In Fig.~\ref{TMD-DGLAP} the transverse momentum distributions obtained for down and charm quarks are shown for \PBset~Set2, with 
$z_M \to 1$, i.e.\ regions $(a+b)$, with (red curve) and without intrinsic-\kt\ distribution applied 
(blue curve - a Gauss distribution with $q_s=0.00001$~\GeV).
We also show the transverse momentum distribution contribution from region $(a)$ alone, i.e.\ for 
$z_M = \zdyn = 1 - q_0/\mu^\prime$,
without intrinsic-\kt\ (corresponding to the magenta curve of Fig.~\ref{PB-DGLAP}). 
The importance of the large $z$-region on the transverse momentum distributions is seen in the comparison with the 
predictions without intrinsic-\kt\ distribution (blue and magenta curves).

The transverse momentum distributions show very clearly the large effect of the choice of $z_M$ for the soft region, while in the perturbative region $\kt > q_0$ the effect becomes smaller with increasing \kt .
Applying such a scale, $z_M=\zdyn = 1 - q_0/\mu^\prime$, removes emissions with $\qt < q_0$ (there are still 
low-\kt\ contributions, which come from adding vectorially all intermediate emissions). However, very soft emissions 
are automatically included with $z_M \to 1$. 

As shown in Fig.~\ref{TMD-DGLAP}, the effect of the intrinsic-\kt\ distribution is much reduced at large scales, but the contribution 
of the region $\zdyn < z < 1$ stays important for small \kt .

It is interesting to observe that the charm density shows essentially no effect of an intrinsic-\kt\  distribution: this is because 
charm is generated dynamically from gluons only, and there is no intrinsic charm density.

\subsection{Transverse momentum distributions of \PBset}
After having discussed  the importance of the soft nonperturbative region to the transverse momentum distribution, 
we turn now to a discussion of the transverse component of the \PBM\ parton distributions of Ref.~\cite{Martinez:2018jxt}, 
which are used for comparison with measurements. 

 In previous investigations on \PZ -boson production at the LHC~\cite{Martinez:2019mwt}, 
 as well as for low DY mass, \mdy ,
and at low  $\sqrt{s}$~\cite{Martinez:2020fzs}, 
it was found that \PBset~Set2 describes the measurements much better, 
while \PBset~Set~1 gives too large a cross section at small DY lepton pair transverse momenta, \ptll.
\begin{figure}[h!tb]
\begin{center} 
\vskip -2cm
\includegraphics[width=0.51\textwidth]{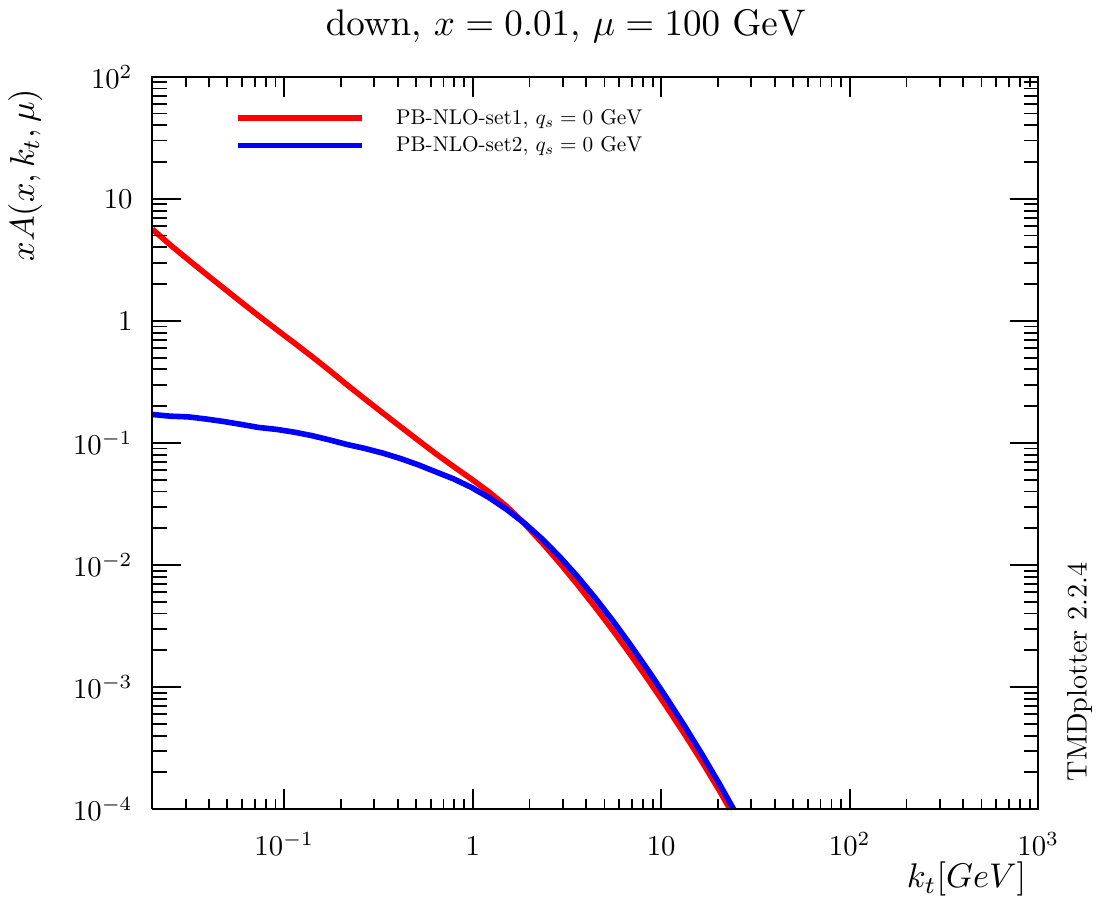} \hskip -0.8cm
\includegraphics[width=0.51\textwidth]{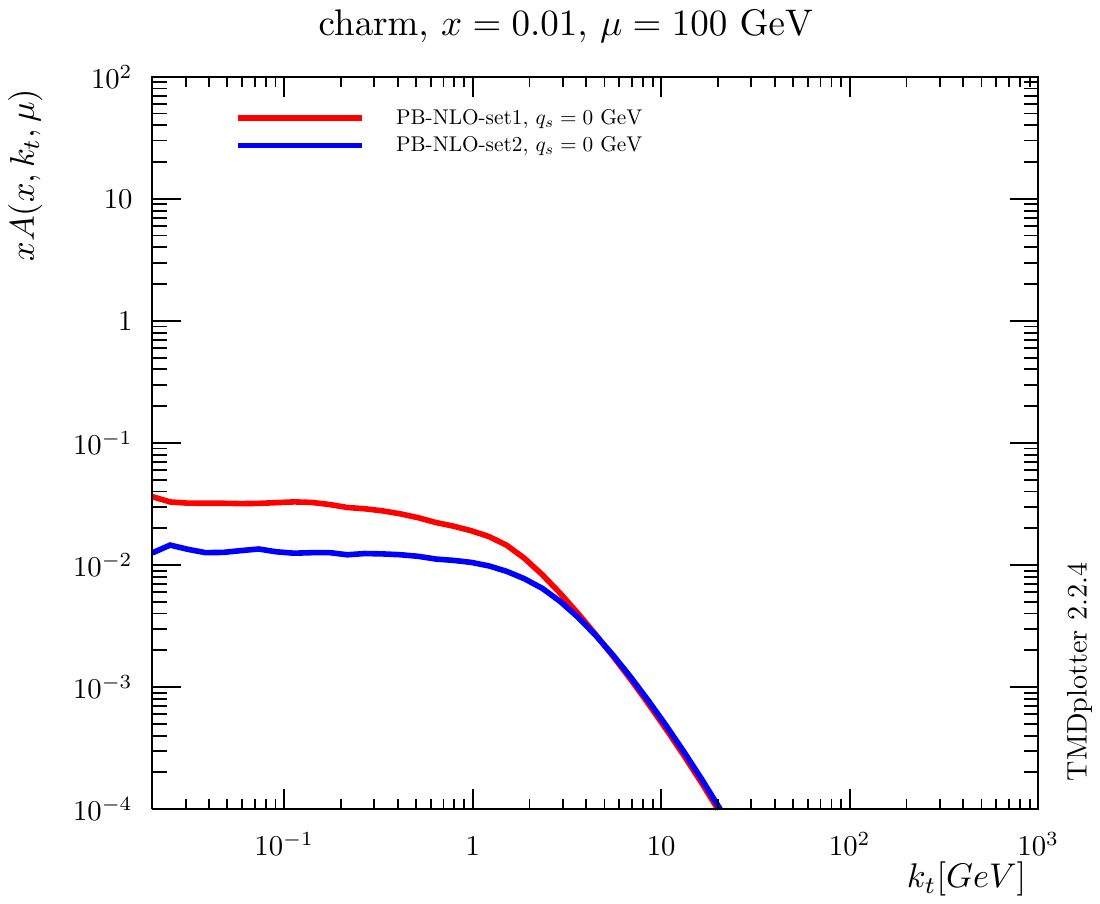} 
\vskip -5.2cm
\includegraphics[width=0.51\textwidth]{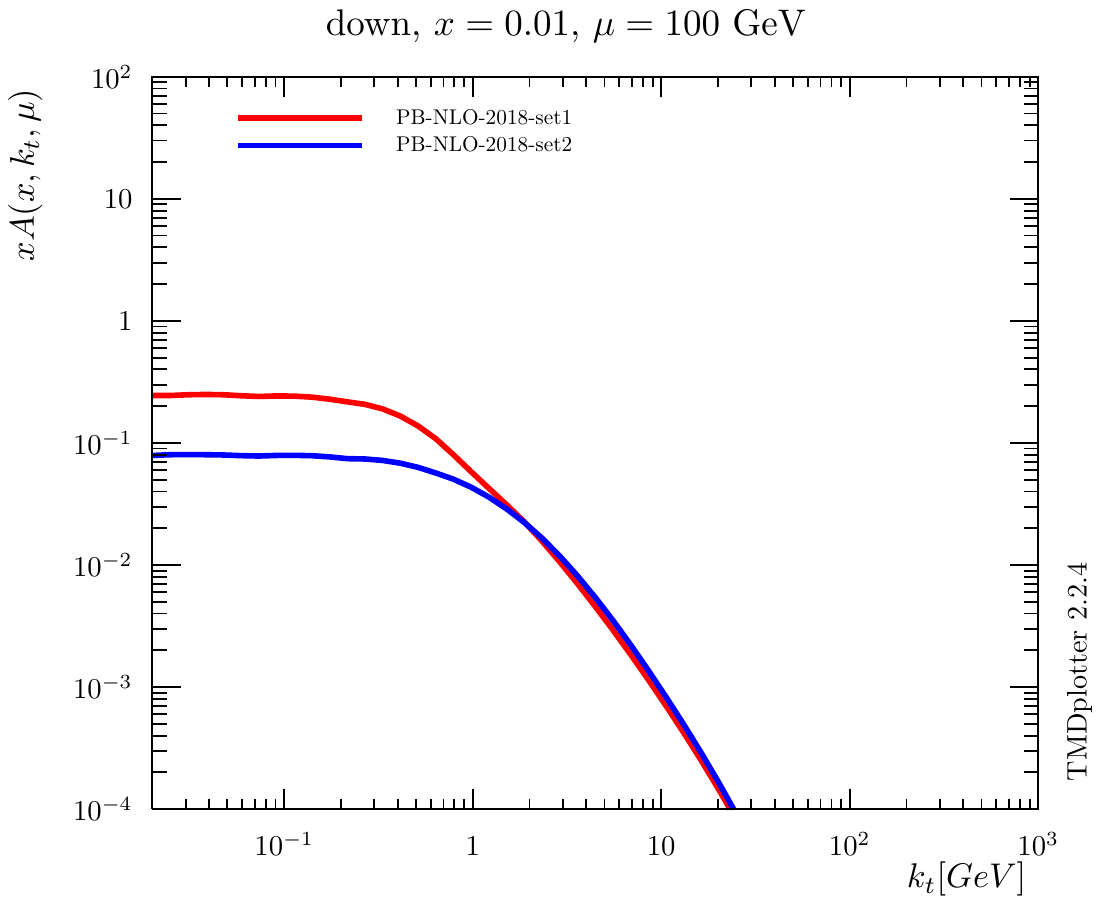} \hskip -0.6cm
\includegraphics[width=0.51\textwidth]{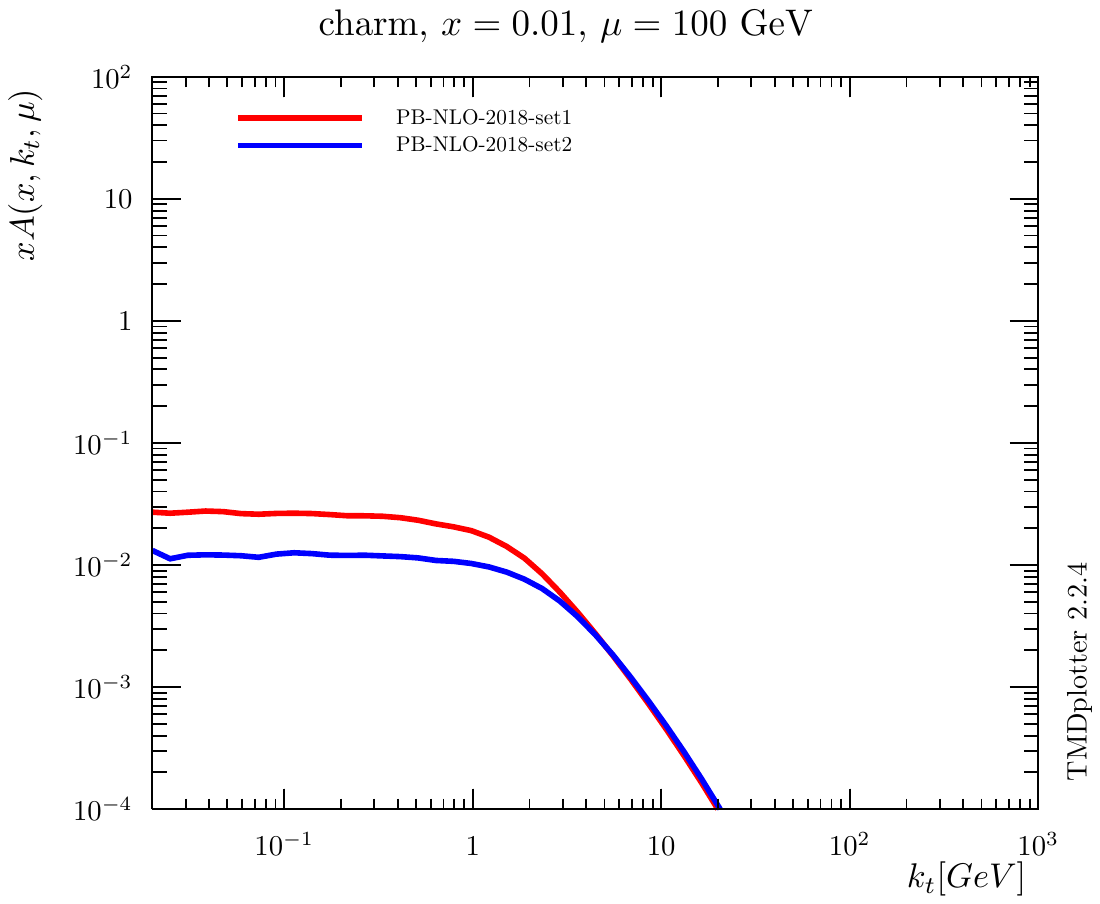} 
\vskip -2.5cm
  \caption{\small 
  TMD parton density distributions for down and charm quarks of the published  \PBset~Set1 (red curve) and \PBset~Set2 (blue curve) \protect\cite{Martinez:2018jxt} as a 
  function of $\kt$ at $\mu=100$~GeV and $x=0.01$.
  In the upper row are shown distributions when no intrinsic-\kt\ distribution is included ($q_s=0.00001$~GeV), and the 
  lower row shows the default distributions with $q_s=0.5$~GeV.
 }
\label{TMD_pdfs}
\end{center}
\end{figure} 

The difference between \PBset~Set1 and Set2, which comes from the choice of renormalization scale (argument in \alphas ), is seen
essentially in the low-\kt\ region, where the nonperturbative Sudakov form factor (region $(b)$), with the integral $z_M \to 1$, 
plays an important role. In Fig.~\ref{TMD_pdfs} (upper row) the distributions for up and charm quarks are shown when no intrinsic-\kt\ 
distribution is included, the lower row shows distributions including the default intrinsic Gauss \kt\ distributions of widths $q_s = 0.5$~GeV. 
It is very interesting to observe that the differences between the sets setting $q_s =0$ or not are very much reduced 
for heavy flavors since they are only generated dynamically 
(since heavy flavors are not present at the starting scale in the VFNS which is applied here).
In principle an intrinsic charm contribution can be included in \PBM\ densities, however, this is not required from inclusive DIS data\cite{Abramowicz:2015mha} used in the fit of \PBset~Set2.

\begin{figure}[h!tb]
 \begin{center} 
  \vskip -1 cm
  \includegraphics[width=0.34\textwidth]{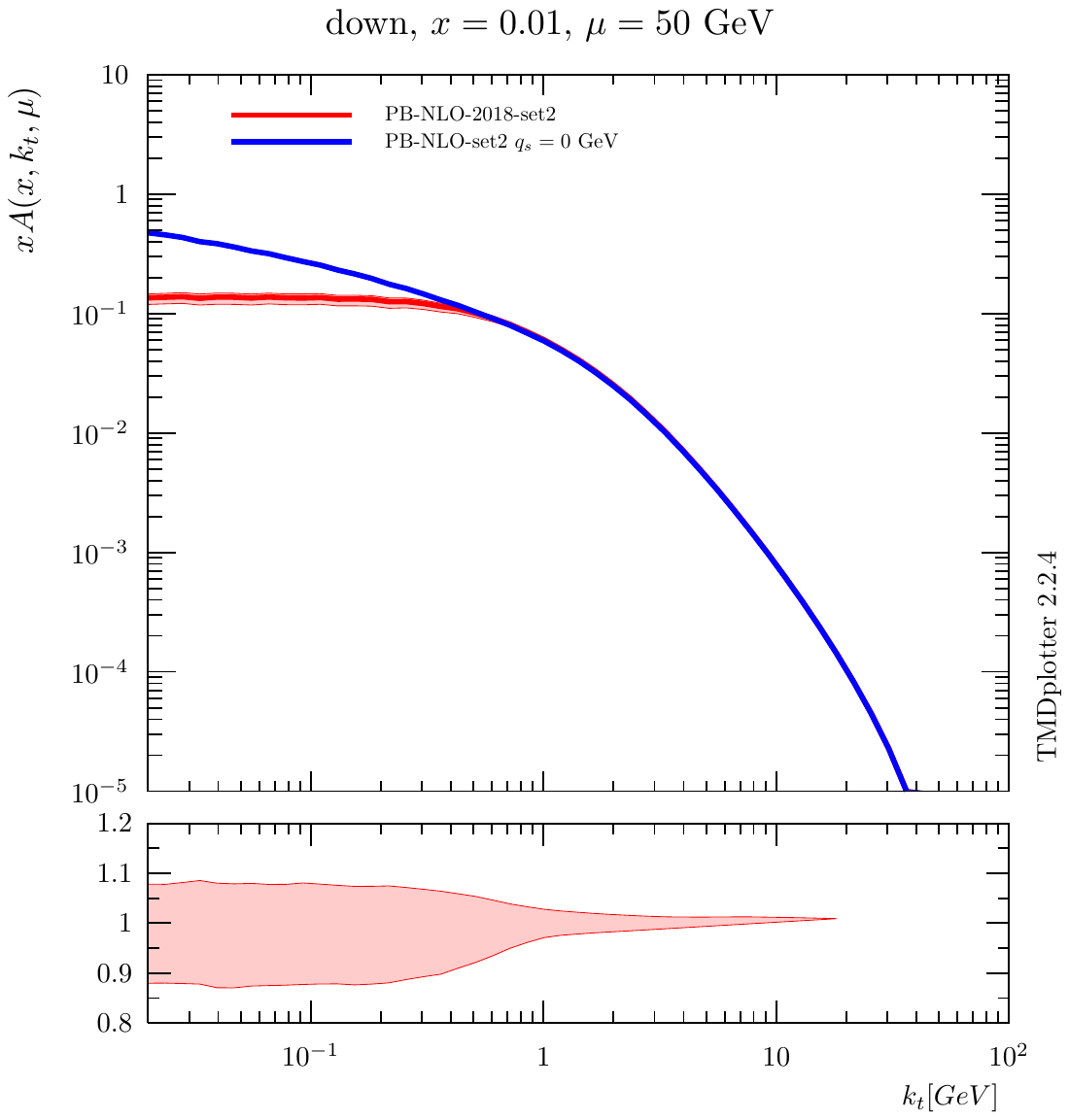} \hskip -0.6cm
  \includegraphics[width=0.34\textwidth]{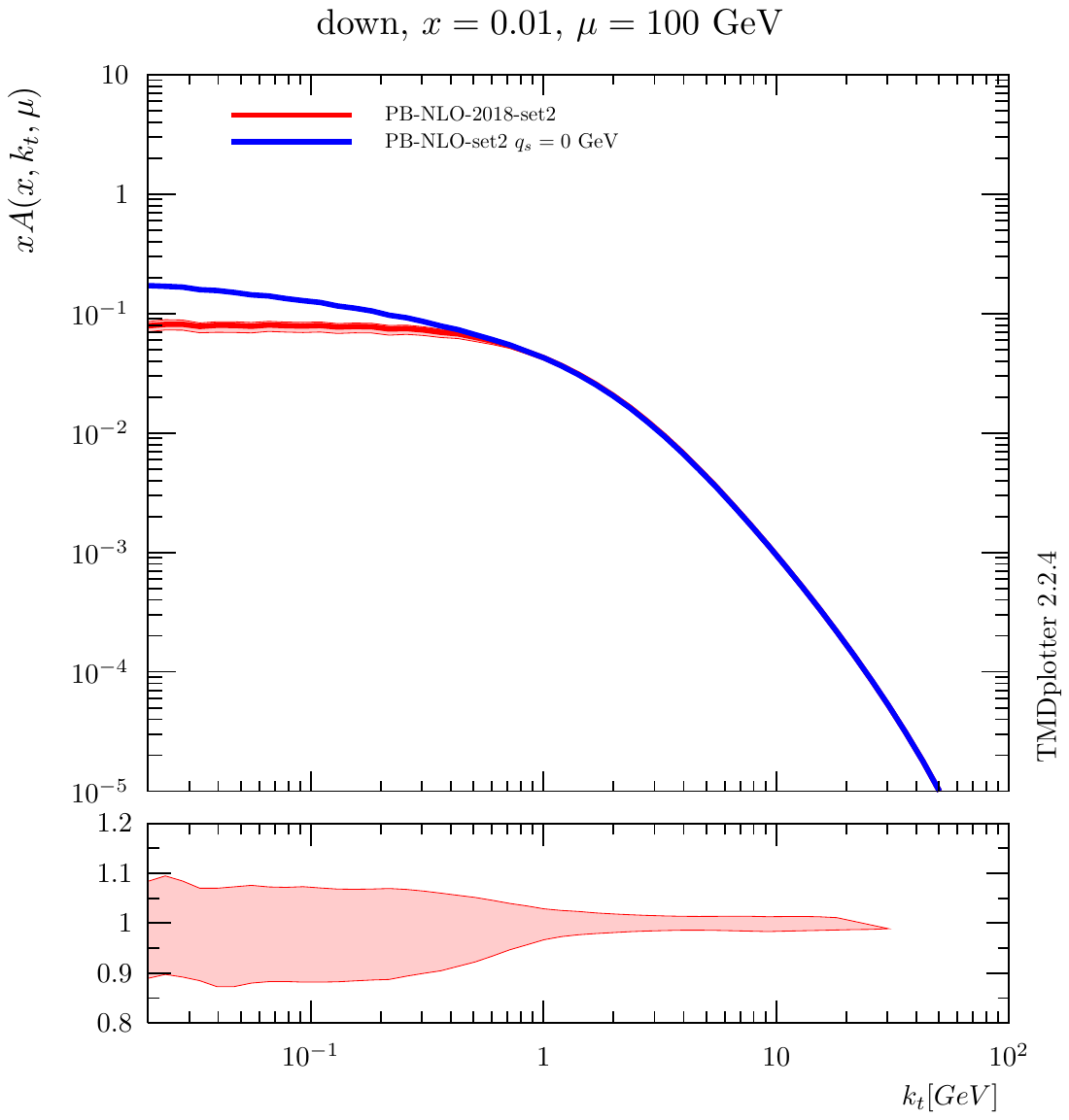} \hskip -0.6cm
  \includegraphics[width=0.34\textwidth]{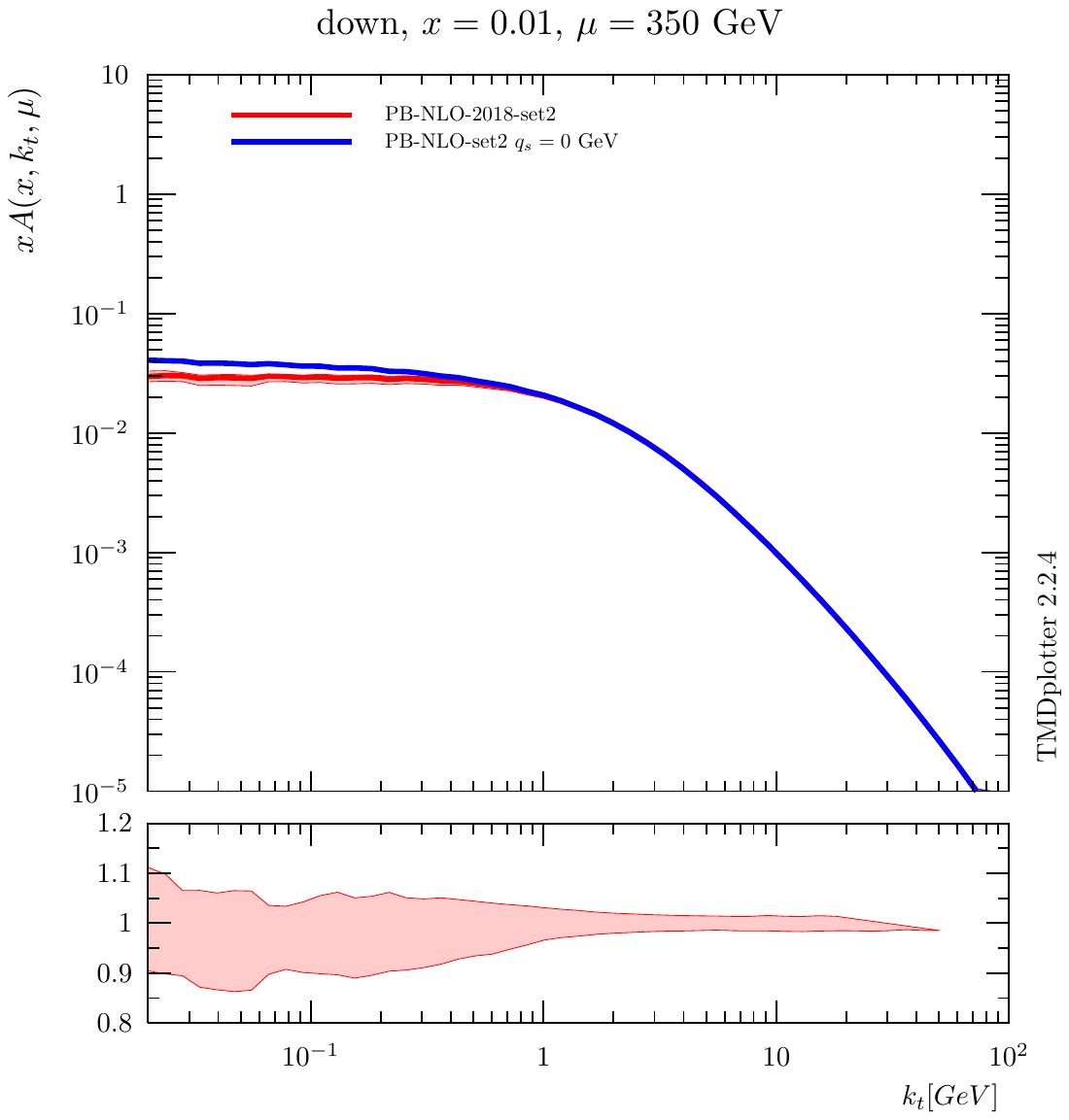} 
  \vskip -1 cm
  \caption{\small TMD parton density distributions for down quarks  of \PBset~Set2 with (red curve) and without (blue curve) 
   intrinsic-\kt\ distribution as a function of $\kt$ at different scales $\mu$ and $x=0.01$.
   The lower panels show the full uncertainty of the TMD PDFs, as obtained from the fits \protect\cite{Martinez:2018jxt}. Shown is the ratio to each central value. The red band shows the uncertainty of \PBset~Set2, the blue line has no uncertainty band.
  }
  \label{TMD_pdfs_scale}
 \end{center}
\end{figure} 

In Fig.~\ref{TMD_pdfs_scale} the transverse momentum distribution for down quarks, with and without  an intrinsic-\kt\ distribution, 
is shown at different scales $\mu$. While at low scales $\mu \sim 50 $~GeV a significant effect of the intrinsic-\kt\ distribution 
is observed for very small \kt , at large scales $\mu \sim 350$~GeV this effect is much reduced. 
This scale dependence will result in a much smaller sensitivity to the intrinsic-\kt\ distribution at high \mdy .

\subsection{Calculation of the DY cross section} \label{sec:xsection}
\begin{tolerant}{2000}
The cross section of DY production is calculated at NLO with \MCatNLO~\cite{Alwall:2014hca}. In the MCatNLO method, 
the collinear and soft contributions of the NLO cross section are subtracted, as they will be later included when parton shower, 
or as in our case, TMD parton densities are applied. As in earlier studies, we use  \cascade 3~\cite{Baranov:2021uol} to include 
TMD parton distributions and parton shower to the MCatNLO calculation (a  detailed investigation of the effect of TMD parton 
distributions and parton showers applied in the \cascade 3 Monte Carlo generator is given in Ref.~\cite{Yang:2022qgk}). 
We use the \herwig6 subtraction terms in MCatNLO, since they are based on the same angular ordering conditions as the \PBM TMD parton 
distribution sets, \PBset~Set1 and \PBset~Set2,  described in the previous section. The validity and consistency using \herwig6 
subtraction terms in MCatNLO together with \PBM\ TMD distributions has been studied in detail in the appendix of Ref.\cite{Yang:2022qgk}.
The predicted cross sections  (labeled as \CAS\ in the following) are calculated using the integrated versions of the NLO parton 
densities \PBset~Set1\ and \PBset~Set2\, together with $\alphas (m_{\PZ})=0.118$ at NLO.
\end{tolerant}

\begin{tolerant}{2000}
The factorization scale $\mu$, used in the calculation of the hard process is set to $\mu =
\frac{1}{2} \sum_i \sqrt{m^2_i +p^2 _{t,i}}$, with the sum running over all final state particles, in
case of DY production over all decay leptons and the final jet. For the generation of transverse
momentum according to the \PBM -TMD distributions, the factorisation scale $\mu$ in the hard process is set to $\mu =
\mdy$,  in the case of a real emission it is set to $\mu= \frac{1}{2} \sum_i \sqrt{m^2_i +p^2
_{t,i}}$. The generated transverse momentum is limited by the matching scale $\mu_{m}=$\verb+SCALUP+
\cite{Baranov:2021uol}.
Since there are no \PBM-fragmentation functions available yet, 
the final state parton shower  in \cascade 3 is generated from \pythia~\cite{Sjostrand:2006za}, including photon radiation of 
the lepton pair.
\end{tolerant}

\begin{figure}[h!tb]
 \begin{center} 
  \includegraphics[width=0.49\textwidth]{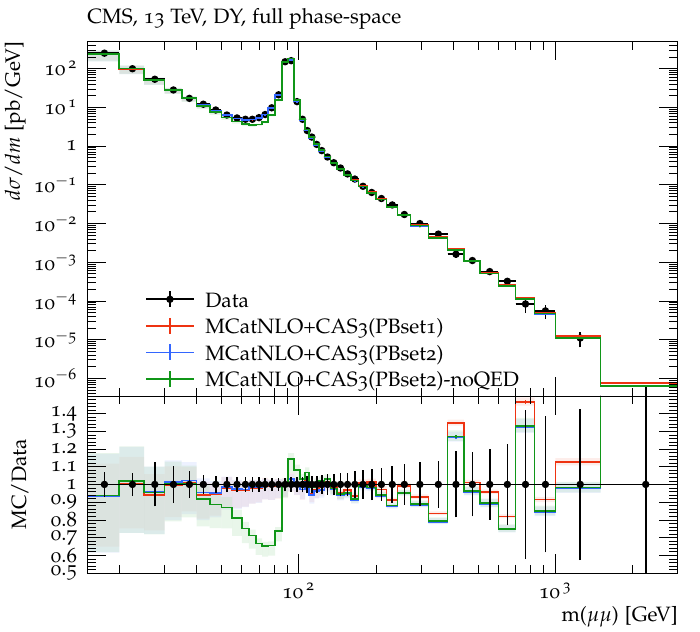}
  \includegraphics[width=0.49\textwidth]{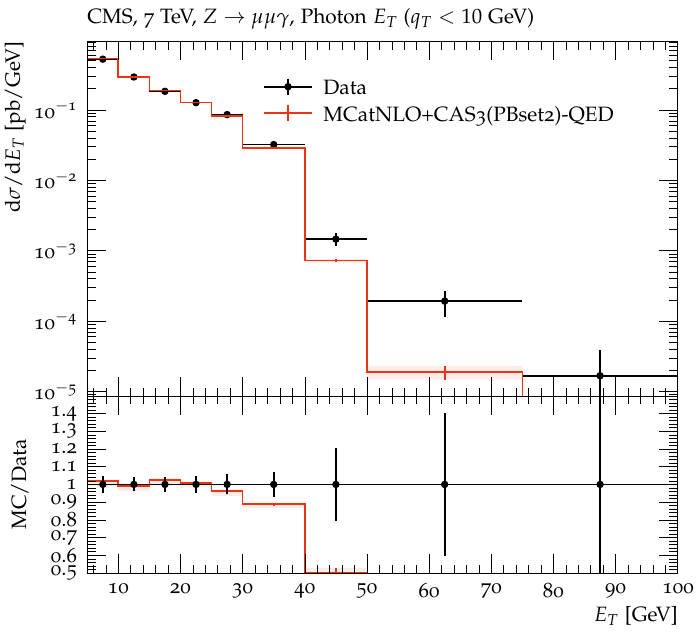}
  \caption{\small {\bf Left}: The mass distribution of DY lepton pairs at 13~TeV \protect\cite{Sirunyan:2018owv} compared to 
   predictions of \CAS\ with \PBset~Set1 (red curve), \PBset~Set2 (blue curve) and without QED corrections (green curve).
  {\bf Right}: The spectrum of photons transverse momentum in $\PZ \to \mu^+ \mu^- \gamma$ at 7~TeV \protect\cite{CMS:2015vap} 
  compared to \CAS\ \PBset~Set2 including QED radiation for a transverse momentum of the DY pair $\ptll < 10$~GeV. 
  The bands show the scale uncertainty.
  }
\label{PB-QED}
\end{center}
\end{figure}

A good description of the final state QED corrections, and in particular the kinematic effect of the real photon radiations, is essential in order to achieve a precise description of the DY transverse momentum. 
Fig.~\ref{PB-QED} (left) shows the DY mass distribution as measured by CMS~\cite{Sirunyan:2018owv} at 13~TeV together with  predictions of \CAS\ \footnote{We use the Rivet package \cite{Buckley:2010ar} for the calculation of the final distributions.}. 
The bands show the scale uncertainty coming from a variation of the renormalization and factorization scale by a factor of two up and down, avoiding the extreme values (7-point variation). 
The DY mass is calculated from the so-called {\it dressed-leptons} (see for example~\cite{Sirunyan:2019bzr,Aad:2014qja}), where photons radiated within a cone of  radius of $R<0.1$ are merged to the lepton before the momenta are calculated. 
We show predictions based on \PBset~Set1 and Set2, and also, for illustration,  when photon radiation is turned off in the final-state shower (labeled as "noQED"). 
A rather good description of the DY mass spectrum over a large range on \mdy\ is obtained both with \PBset~Set1 and Set2. 
Only at $\mdy$ greater than a few hundred GeV the predictions tend to become smaller than the measurement (while still within the uncertainties). 
However, this is the region where the partonic $x$ becomes large and not well constrained by the fit to HERA 
data~\cite{Abramowicz:2015mha} used for the \PBset\ TMD extraction \cite{Martinez:2018jxt}. 
In the region of \mdy below the \PZ-pole, one can observe the importance of QED corrections.
In Fig.~\ref{PB-QED} (right) we show the photon transverse momentum spectrum in \PZ -production as measured by
CMS~\cite{CMS:2015vap} at 7~TeV in comparison with \CAS\ including QED radiation. The photon spectrum is well described at low $E_T
< 40$~GeV, while the high $E_T$ spectrum predicted by the parton shower falls below the measurement, since the precision of parton showers
are limited for the high \pt\ region.

\section{The transverse momentum spectrum of DY lepton pairs } \label{sec:pt}
\begin{figure}[h!tb]
 \begin{center} 
  \includegraphics[width=0.32\textwidth]{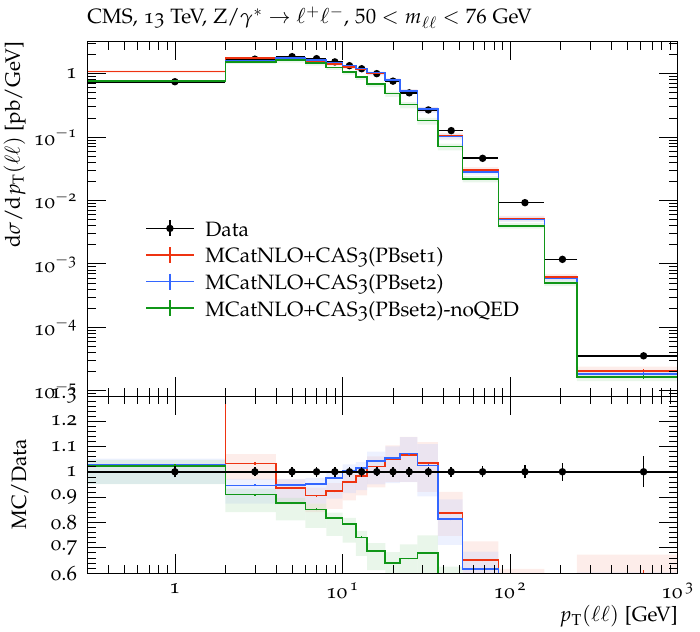}
  \includegraphics[width=0.32\textwidth]{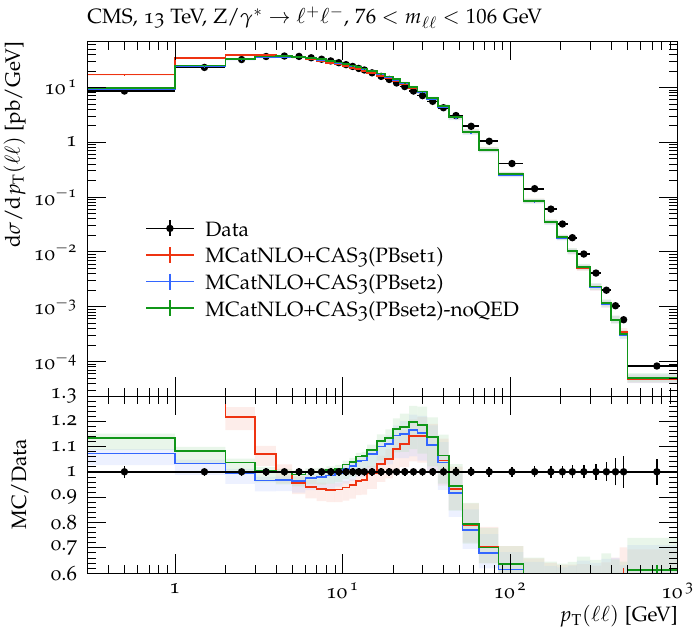}
  \includegraphics[width=0.32\textwidth]{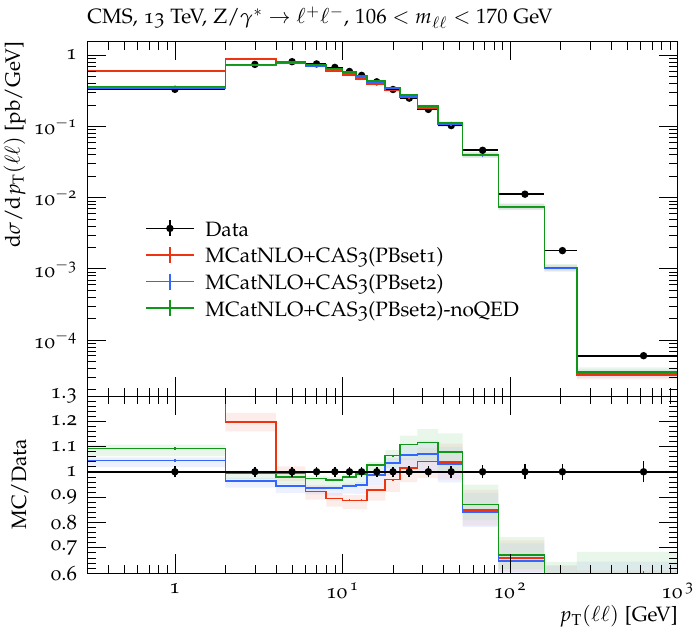}
  \caption{\small The \ptll\ dependent DY cross section for different \mdy\ regions as measured by CMS~\protect\cite{CMS:2022ubq} 
   compared to \CAS\ predictions based on \PBset~Set 1 (red curve) and Set 2 (blue curve). 
   Also shown are predictions without the inclusion of final state QED radiation from the leptons (green curve). 
   The band shows the 7-point variation of the renormalization and factorization scales.
  }
  \label{CAS_CMS_WideMass}
 \end{center}
\end{figure}

The transverse momentum spectrum of DY pairs at $\sqrt{s}=13$~TeV has been measured for a wide  \mdy\ range  by CMS~\cite{CMS:2022ubq}. 
We use this measurement for comparison with predictions of \CAS\ based on \PBset~Set1 and \PBset~Set2, as shown in Fig.~\ref{CAS_CMS_WideMass}. 
As already observed in previous investigations 
 \cite{Yang:2022qgk,Abdulhamid:2021xtt,BermudezMartinez:2020tys,Martinez:2019mwt}, 
the \PBset~Set1 gives too high a contribution at small transverse momenta \ptll , while \PBset~Set2 describes the measurements 
rather well, without any further adjustment of parameters\footnote{The predictions shown here are slightly different compared 
to the predictions in \protect\cite{CMS:2022ubq} because we use here a lower minimum \kt\ cut and because of a bug in the 
treatment of QED radiation in Rivet, corrected in version~3.1.8}, underlining 
the role of evaluating the strong coupling at the transverse momentum scale. 
In order to illustrate the importance of QED corrections, we show in addition a prediction based on \PBset~Set2 without 
including QED final state radiation (labeled noQED). 
Especially in the low \mdy\ region, the inclusion of QED radiation is essential, not only changing the total cross section but 
rather strongly modifying the shape of the transverse momentum distribution \ptll . 
All calculations predict too low a cross section at large transverse momentum due to missing higher-order contributions in the matrix 
element.
In Refs.~\cite{BermudezMartinez:2022bpj,BermudezMartinez:2021lxz,Martinez:2021dwx} it is shown explicitly that including higher 
orders in the matrix element through the TMD multi-jet merging technique gives an excellent description even for largest \ptll . For all further distributions, we restrict 
the investigations to \ptll\ below the peak region (i.e.\ \ptll $\lesssim 8$ GeV).

\subsection{Influence of the intrinsic-\boldmath\kt\ distribution on DY transverse momentum distributions}
Given the rather successful description of the DY \ptll -spectrum with \CAS\ using  \PBset~Set 2 in the low \ptll -region, we investigate
below the importance of the intrinsic-\kt\ distribution. 
In \PBset\ the intrinsic-\kt\ distribution is parameterized as a Gauss distribution with zero mean and a width 
$\sigma^2 =q_s^2/2$ \cite{Martinez:2018jxt} (see Eq.~(\ref{TMD_A0})), 
where $q_s$ was fixed by default at $q_s = 0.5 $ \GeV .

\begin{figure}[h!tb]
\begin{center} 
\includegraphics[width=0.60\textwidth]{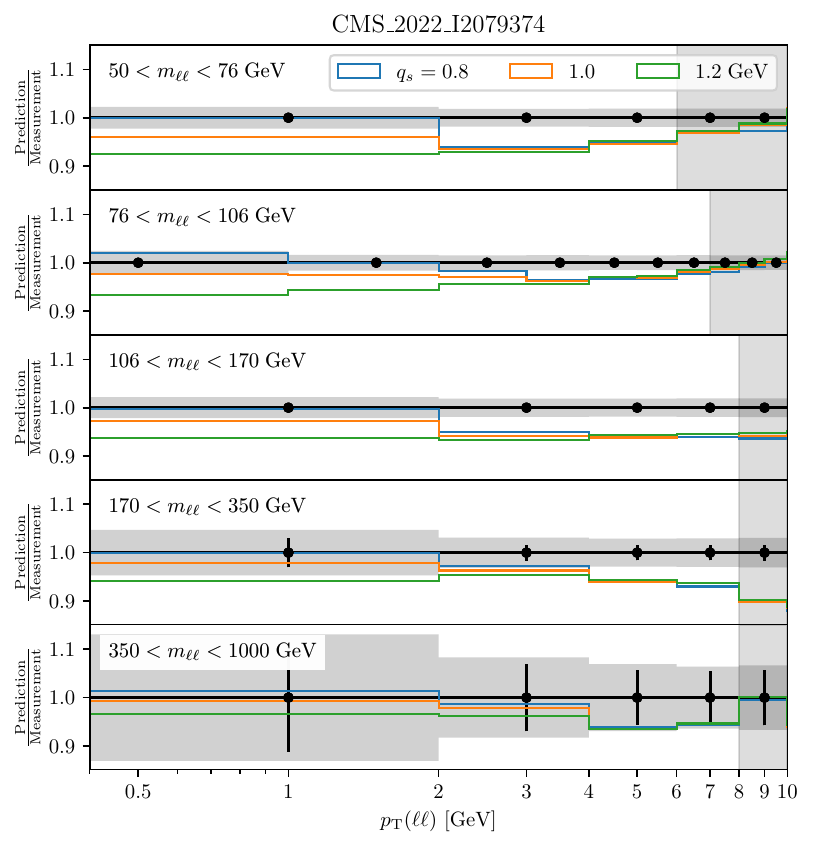}
  \caption{\small Drell-Yan cross section ratios of \CAS\ predictions for different  $q_s$ values over CMS measurement~\protect\cite{CMS:2022ubq} as a function of \ptll\ for different \mdy\ regions. Only the lowest \ptll\ values are shown. The points error bar show the statistical uncertainties and the gray bands the total experimental uncertainties. The gray area at the highest \ptll\ values show the maximal values included in the fit described in section \ref{qsFit}.
  }
\label{CAS_set2-noIntrisicKt}
\end{center}
\end{figure}

In order to illustrate the sensitivity range of the intrinsic-\kt\ distribution, we show in Fig.~\ref{CAS_set2-noIntrisicKt} the \CAS\ predictions for the low \ptll -spectrum of DY production at different DY masses \mdy\ 
for different intrinsic-\kt\ distribution (with different $q_s$ parameter values) compared to the CMS measurement~\protect\cite{CMS:2022ubq}. 
We observe that sensitivity to intrinsic-\kt\ is more pronounced at small \ptll\ values. This sensitivity decreases
with increasing mass, as expected from Fig.~\ref{TMD_pdfs_scale}

\begin{tolerant}{2000}
In the following we describe a determination of the Gaussian width $q_s$ for different DY masses, \mdy , at different $\sqrt{s}$. 
The prediction is obtained from a calculation of \CAS\ using TMD distributions obtained with the \PBset~Set2 parameters for the collinear distribution, but with different $q_s$ values.
We scan for each \mdy -bin  $q_s$  in steps of $0.1$ to $0.3$~GeV in the range $q_s = 0.1, \dots , 2.0$~GeV. 
At higher DY transverse momenta, higher order contributions have to be taken into account (a study using multijet merging is given in Refs.~\cite{BermudezMartinez:2022bpj,BermudezMartinez:2021lxz,Martinez:2021dwx}).
\end{tolerant}

\subsection{Fit of the Gauss width \boldmath$q_s$ in pp at $\sqrt{s}=13$ TeV} \label{qsFit}

The transverse momentum distribution of DY leptons has been measured by the CMS collaboration~\cite{CMS:2022ubq}. 
This is the basic measurement for the determination of the intrinsic-\kt\ parameter $q_s$, since it covers a wide \mdy -range with
high precision and that a detailed uncertainty breakdown, discussed in subsection~\ref{CMS_DY}, is provided. 
The measurements of \PZ -production obtained from  
LHCb~\cite{LHCb:2021huf} are discussed in subsection~\ref{LHCb}, while measurements at lower center-of-mass energies 
are shown in subsection~\ref{lowCM}.

\subsubsection{DY production over a wide DY mass range \label{CMS_DY}}

The CMS collaboration has measured Drell-Yan production at 13~TeV \cite{CMS:2022ubq} covering a range of DY mass 
$\mdy=[50, 76, 106, 170, 350, 1000]$~GeV. 
The measurement is provided with a detailed uncertainty breakdown, corresponding to a complete treatment of experimental
uncertainties including correlations between bins of the measurement for each uncertainty source separately.
Note that we use the fully detailed breakdown of the experimental uncertainties provided on the CMS
public website \footnote{The corresponding HEPdata records only contain summarised information}.

In order to determine the intrinsic-\kt\ we vary the $q_s$ parameter and calculate a $\chi^2$ to quantify the model agreement with the 
measurement.
We evaluate the following expression\footnote{The code used with the full covariance matrix is available in Ref.~\cite{Chi2LMoureaux}, an earlier version to be used directly with Rivet is in Ref.~\cite{lhcew-LCorpe}.},
\begin{equation}
\chi^2 = \sum_{i,k} (m_i-\mu_i) C_{ik}^{-1}(m_k-\mu_k),
\end{equation}
with $m_i$ being the measurement and $\mu_i$ being the prediction for data point $i$. The
covariance matrix $C_{ik}$ is decomposed into a component describing the uncertainty in the
measurement, $C_{ik}^\text{meas.}$, and the statistical and scale uncertainties in
the prediction,
\begin{equation}
C_{ik} = C_{ik}^\text{meas.} + C_{ik}^\text{model-stat.} + C_{ik}^\text{scale}.
\label{chi2corr}
\end{equation}
The covariance matrix of the measurement is taken directly from the supplementary material
provided by CMS.
The statistical uncertainty in the prediction, arising from the use of a Monte Carlo
simulation, is accounted for as a small diagonal contribution without correlations between
bins,
\begin{equation}
C_{ik}^\text{model-stat.} = \sigma^2_{i,\text{stat.}} \; \delta_{ik},
\end{equation}
where $\sigma_{i,\text{stat.}}$ is the bin-by-bin statistical uncertainty.
We also treat, for the first time, the scale uncertainties of the theoretical prediction as
a correlated uncertainty, for a given \mdy\ range, allowing for a global shift of all bins together within the band
defined by the symmetrized envelope of the scale uncertainties.
This contribution to the covariance matrix is constructed as follows,
\begin{equation}\label{eq:scale-cov-matrix}
C_{ik}^\text{scale} = \sigma_{i,\text{scale}} \; \sigma_{k,\text{scale}},
\end{equation}
where $\sigma_{i,\text{scale}}$ encodes the scale variation for each bin.

\begin{figure}
    \centering
    \includegraphics[width=0.60\textwidth]{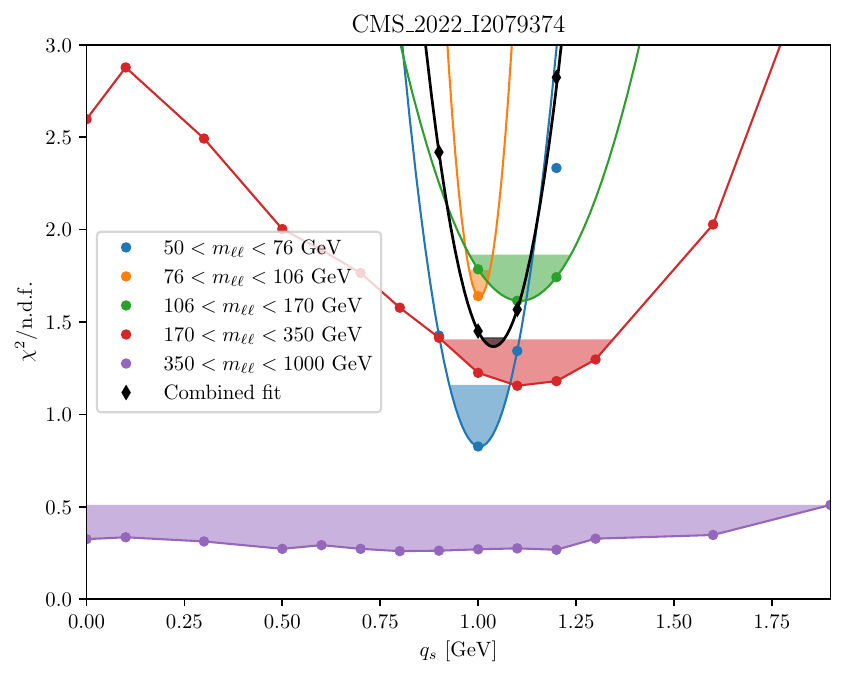}
    \caption{%
      The reduced $\chi^2/\text{n.d.f}$ distribution as a function of $q_s$ for 
      different \mdy regions obtained from a comparison of the \CAS\ prediction with the measurement by CMS~\protect\cite{CMS:2022ubq}. 
      The points represent the obtained $\chi^2$ values. The lines represent the curves used for the uncertainty estimate (see text), which is 
      materialized by the shaded areas.
    }
    \label{fig:chi2ScanCMS-13TeV}
\end{figure}

We first extract independent values of $q_s$ for each invariant mass region considered in the
measurement, considering only the region most sensitive to $q_s$, $\ptll<8\:\GeV$. We
reduce this range further in the first two regions to $\ptll<6\:\GeV$ for
$50<\mdy<76\:\GeV$ and $\ptll<7\:\GeV$ for $76<\mdy<106\:\GeV$ to stay in the region of sensitivity 
and not be biased by missing higher orders in the predictions affecting high \ptll\ shape. 
The obtained  $\chi^2/\text{n.d.f}$ (reduced $\chi^2)$)  values are shown in Fig.~\ref{fig:chi2ScanCMS-13TeV} as a function of $q_s$.

\begin{figure}
    \centering
    \includegraphics[width=0.65\textwidth]{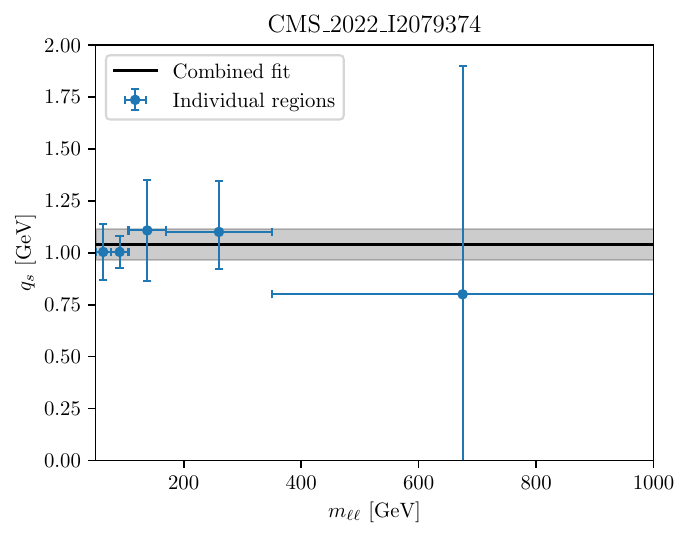}
    \caption{%
        The values of $q_s$ in each \mdy -bin as obtained from 
        Ref.~\protect\cite{CMS:2022ubq}. Indicated is also the combined fit value of $q_s$.
    }
    \label{fig:qs_vrs_mdy-13TeV}
\end{figure}

Within each region, we consider the value of $q_s$ for which the smallest $\chi^2$ is
obtained as our ``best fit'' value. We construct a one-sigma confidence region as the set
of all $q_s$ values for which $\chi^2(q_s) < \min(\chi^2) + 1$. When possible, this region
is determined graphically using a linear interpolation between scan points. When the 
minimum is too narrow for a reliable determination of the uncertainty using this method,
we use instead a quadratic interpolation between the lowest three points and add an
uncertainty equal to one half-bin-width (0.05\:\GeV) in quadrature. In addition, we
include an uncertainty derived by repeating the procedure with modified fit boundaries.
The values obtained using this method are listed in Table~\ref{table:qs_cms} and a
comparison is shown in Fig.~\ref{fig:qs_vrs_mdy-13TeV}.

\begin{table}
    \centering
    \begin{tabular}{|c|cc|l|}
        \hline
        \mdy region & Best $\chi^2$ & $\text{n.d.f}$ & Best fit $q_s$ [GeV] \\
        \hline
        \phantom{00}50--76\:\GeV\phantom{00}
            & 2.45 & 3
            & $1.00\pm0.08(\text{data})\pm0.05(\text{scan})\pm0.1(\text{bins})$ \\
        \phantom{00}76--106\:\GeV\phantom{0}
            & 11.4 & 7
            & $1.03\pm0.03(\text{data})\pm0.05(\text{scan})\pm0.05(\text{bins})$ \\
        \phantom{0}106--170\:\GeV\phantom{0}
            & 6.46 & 4
            & $1.11\pm0.13(\text{data})\pm0.05(\text{scan})\pm0.2(\text{bins})$ \\
        \phantom{0}170--350\:\GeV\phantom{0}
            & 4.62 & 4
            & $1.1^{+0.24}_{-0.18}(\text{data})$ \\
        350--1000\:\GeV
            & 1.04 & 4
            & $<1.9$ \\
        \hline
    \end{tabular}
    \caption{%
        Results of the fit on individual \mdy intervals for the CMS
        measurement~\protect\cite{CMS:2022ubq}. The ``data'' uncertainty is the one 
        estimated using $\min(\chi^2)+1$, the ``scan'' uncertainty accounts for the step
        size of the $q_s$ scan, and the ``bins'' uncertainty is derived by varying the
        number of bins included in the fit. The number of bins used in the fit gives $\text{n.d.f}$.
    }
    \label{table:qs_cms}
\end{table}

The values derived from each \mdy\ interval are compatible with each other. The most precise
determination is obtained from the \PZ{} peak region, $76<\mdy<106\:\GeV$, followed by the
 regions around it. The sensitivity at high mass suffers from larger statistical
uncertainties in the measurement.
 This independence of the intrinsic-\kt\ with the DY mass contrasts with the need to tune the Parton Shower parameters 
for different masses in standard Monte Carlo events generators (see \cite{CMS:2022ubq} - Fig.~6 for a data comparison 
with \MCatNLO\ interfaced with \pythia\ Parton Shower).

Having obtained compatible results, we proceed to deriving a combined fit by calculating a
joint $\chi^2$ including the considered bins in all mass ranges. For this, we construct a
new covariance matrix $C_{ik}^\text{comb.}$ as a sum over the 650 uncertainty sources
included in the detailed breakdown. We consider that each systematic uncertainty is fully
correlated between \mdy regions and construct their covariance matrices in the same way as
in Eq.~\eqref{eq:scale-cov-matrix}. The statistical uncertainties (data and Monte Carlo) in
the measurement feature nontrivial correlations due to the use of unfolding but are
independent in each \mdy region, and therefore we construct a block-diagonal matrix from
the covariance matrices in each \mdy region. The statistical uncertainty in the prediction
is diagonal. We consider that the uncertainties in the QCD scales are not correlated between 
\mdy\ regions and use a block-diagonal matrix.

The $\chi^2$ values obtained using the combined covariance matrix are shown in
Fig.~\ref{fig:chi2ScanCMS-13TeV}. The best fit value, extracted in the same way as for
separate regions, is,
\[
    q_s = 1.04\pm0.03(\text{data})\pm0.05(\text{scan})\pm0.05(\text{binning})\:\GeV.
\]
This value and its uncertainty are shown as a black line and shaded area on
Fig.~\ref{fig:qs_vrs_mdy-13TeV} for comparison with the individual \mdy bins.
A cross-check has been performed by interpolating the prediction for each bin between $q_s$ values
and searching for the minimum of the $\chi^2$ distribution using a finer $q_s$ grid. It returned values within the uncertainties quoted above. The TMD distributions including the new $q_s$ value are available in TMDlib and TMDplotter~\cite{Hautmann:2014kza,Abdulov:2021ivr}.

To make consistency checks of the obtained value of $q_s$ and to examine possible trends of its dependence on DY mass 
and centre-of mass energy, the DY measurements at high rapidity and lower collision energies have been analysed. 
Since for these measurements no full error breakdown are available, we treat all uncertainties as being uncorrelated 
and do not include systematic uncertainty coming from the scale variation in the theoretical calculation.

\subsubsection{Z production  at high rapidities at 13 \TeV }  \label{LHCb}
The LHCb collaboration~\cite{LHCb:2021huf} has measured \PZ -production at $\sqrt{s} = 13$~TeV in the forward region, covering a rapdity range of $2 < |y| < 4.5 $.  
\begin{figure}[h!tb]
\begin{center} 
\includegraphics[width=0.55\textwidth]{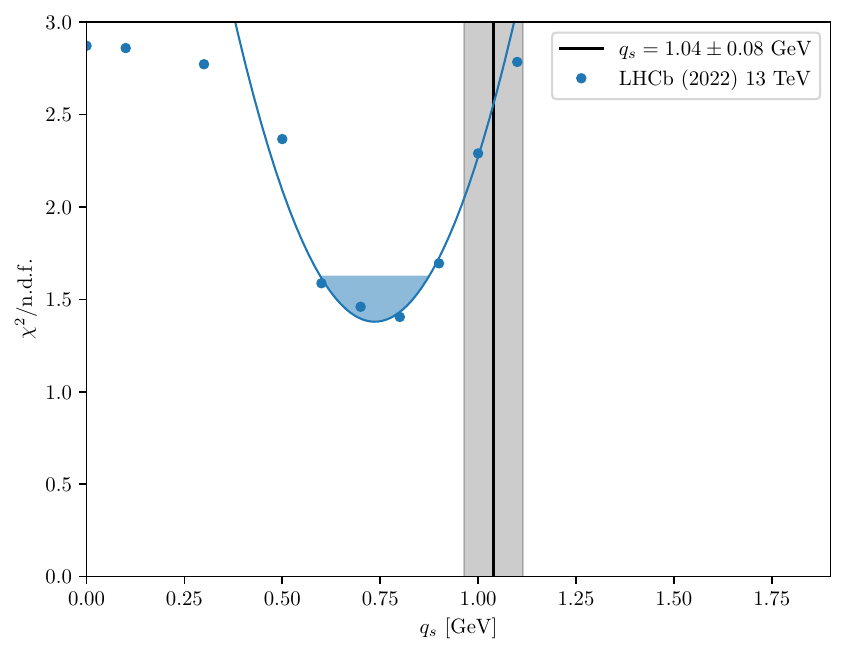}
\caption{\small  The reduced $\chi^2/\text{n.d.f}$ distribution as a function of $q_s$ summed over all rapidity regions obtained from a comparison of the \CAS\ prediction with the measurement by LHCb~\protect\cite{LHCb:2021huf}. The shaded area 
corresponds to $\chi^2+1$. 
The best fit value is $q_s = 0.74 \pm 0.15$~\GeV.  The value of $q_s = 1.04 \pm 0.08$~\GeV\ as obtained from the measurements in Ref.~\protect\cite{CMS:2022ubq} is indicated by a black vertical line.
 }
\label{chi2ScanLHCb-13TeV}
\end{center}
\end{figure} 
The $\chi^2$ distribution is shown in Fig.~\ref{chi2ScanLHCb-13TeV} summed over the rapidity range of the DY lepton pair as a function of $q_s$. 
A minimum is obtained for $q_s=0.74 \pm 0.15 $~\GeV , 
where the uncertainty comes from a variation of $\chi^2$ by one unit and from the step size of the $q_s$ scan.

\subsection{The Gauss widths $q_s$ from lower center of mass energies\label{lowCM}}
\begin{figure}[!htb]
\centering
\minipage{0.50\textwidth}
  \includegraphics[width=\linewidth]{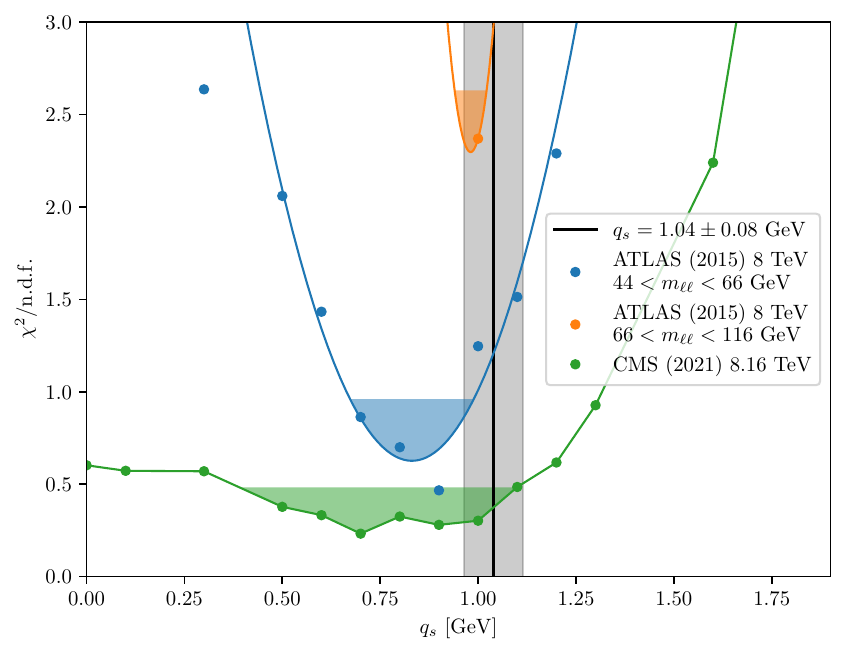}
\endminipage\hfill
\minipage{0.50\textwidth}
  \includegraphics[width=\linewidth]{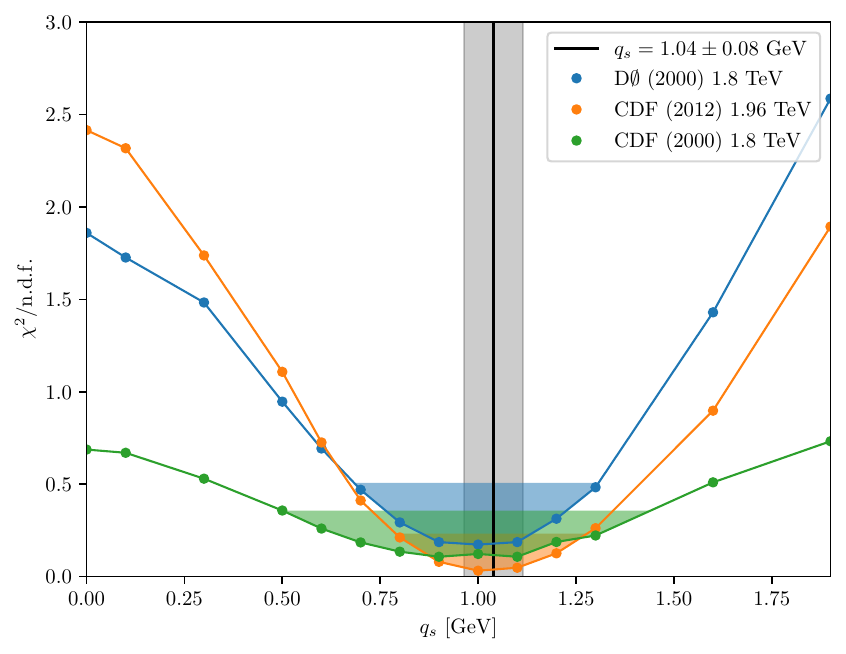}
\endminipage\hfill
\minipage{0.50\textwidth}
 \includegraphics[width=\linewidth]{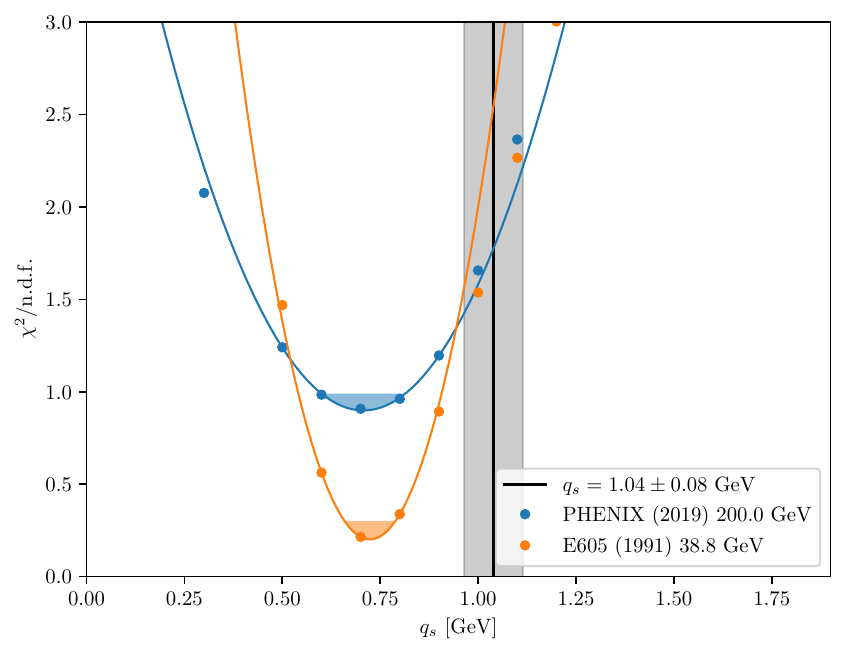}
\endminipage
\caption {The reduced $\chi^2/\text{n.d.f.}$ distribution as a function of $q_s$ obtained from a comparison of the 
  \CAS\ \PBset~Set2 prediction with the measurements at lower center-of-mass energies. 
  The colored shaded band shows the $\chi^2$ variation of one unit for each data set.
  The value of $q_s =1.04 \pm 0.08$~\GeV\ is shown as the grey band. 
 {\bf Left-top}: ATLAS measurement in 2 mass bins that we analysed at $\sqrt{s}= 8$ \TeV\ ($\text{n.d.f.} = 4$ for each mass bin) \protect\cite{Aad:2015auj} 
and CMS in pPb at $\sqrt{s}=~8.1$~\TeV\ \protect\cite{CMS:2021ynu}.
 {\bf Right-top}: Tevatron measurements - D0 at $\sqrt{s}= 1.8$ \TeV\ ($\text{n.d.f.} = 4$) \protect\cite{D0:1999jba}, CDF at $\sqrt{s}= 1.8$ \TeV\ ($\text{n.d.f.} = 5$) \protect\cite{CDF:1999bpw} and $\sqrt{s}= 1.96$ \TeV\ ($\text{n.d.f.} = 6$) \protect\cite{CDF:2012brb} 
 {\bf Bottom}: Measurements at lower energies - PHENIX at $\sqrt{s}= 200$ \GeV\ ($\text{n.d.f.} = 12$) \protect\cite{Aidala:2018ajl}
 and E605 at  $\sqrt{s}= 38.8$ \GeV\ ($\text{n.d.f.} = 11$) \protect\cite{Moreno:1990sf}}.
\label{fig:chiATLASD0}
\end{figure}

The ATLAS collaboration has measured the production of DY from \Pp\!\!\Pp\ collisions at $\sqrt{s} = ~8$~\TeV\ in several DY mass bins, 
out of which only the two with $44 < \mdy < 66$ \GeV\ and $66 < \mdy < 116$ \GeV\ are relevant for $\ptll < 10 $ \GeV\ \cite{Aad:2015auj}. 
 In Fig.~\ref{fig:chiATLASD0} we show the $\chi^2/\text{n.d.f}$ as a function of $q_s$ obtained from these two mass bins ($\text{n.d.f}=8$).

The Tevatron experiments D0~\cite{D0:1999jba} and CDF have measured transverse momenta of DY lepton pairs created 
in $p\bar{p}$ collisions at lower center-of-mass energies (1.8 TeV ~\cite{CDF:1999bpw} and 1.96 TeV \protect\cite{CDF:2012brb}). The PHENIX collaboration measured DY production at $\sqrt{s}=200$~\GeV~\cite{Aidala:2018ajl},  and E605~\cite{Moreno:1990sf} at $\sqrt{s}=38.8$~\GeV .  
The Drell-Yan differential cross section in \ptll\ has also been measured in pPb data at $\sqrt{s} = ~8.1$~\TeV\ by CMS~\protect\cite{CMS:2021ynu}.
Figure~\ref{fig:chiATLASD0} shows the impact that the $q_s$ choice has on $\chi^2/\text{n.d.f}$ for these different measurements.

\subsection{Consistency between determinations of intrinsic \boldmath\kt\ width}
A global fit of $q_s$ is obtained by calculating $\chi^2$ for different measurements, as shown in Table~\ref{table:Saras_table},
including the corresponding center-of-mass energies, collision types and the number of fitted data points, 
resulting in a total of $81$ data points. 

The impact of intrinsic-\kt\ distribution at lower collision energies has  been analyzed using the entire range of \ptll , while at higher center-of-mass energies we investigate up to the peak region in the transverse momentum distribution.  

\begin{table}[ht]
\centering
\resizebox{0.7\textwidth}{!}{\begin{tabular}{|c|c|c|c|}
  \hline
Analysis & $\sqrt{s}$ & Collision types & \text{n.d.f}\\[0.5ex] 
 \hline\hline
CMS\_2022\_I2079374~\protect\cite{CMS:2022ubq} & 13 TeV & pp & 25\\
LHCb\_2022\_I1990313~\protect\cite{LHCb:2021huf} & 13 TeV & pp & 5\\
CMS\_2021\_I1849180~\protect\cite{CMS:2021ynu} & 8.1 TeV & pPb& 5\\
ATLAS\_2015\_I1408516~\protect\cite{Aad:2015auj} & 8 TeV & pp  & 8\\
CDF\_2012\_I1124333~\protect\cite{CDF:2012brb} & 1.96 TeV &  $\rm{p}\bar{\rm{p}}$ &  6\\
CDF\_2000\_S4155203~\protect\cite{CDF:1999bpw} & 1.8 TeV  & $\rm{p}\bar{\rm{p}}$ &  5\\
D0\_2000\_I503361~\protect\cite{D0:1999jba}& 1.8 TeV &  $\rm{p}\bar{\rm{p}}$ &  4\\
PHENIX\_2019\_I1672015~\protect\cite{Aidala:2018ajl} & 200 GeV &  $\rm{p}\bar{\rm{p}}$ & 12\\
E605\_1991\_I302822~\protect\cite{Moreno:1990sf}& 38.8 GeV & pp & 11\\
\hline \hline
Total & & & 81\\
\hline
\end{tabular}}
  \caption{All data sets with the corresponding center-of-mass energies, collision types and the number of degrees of 
  freedom used for the global fit of $q_s$.} 
 \label{table:Saras_table}
\end{table}

\begin{figure}[h!tb]
\begin{center} 
\includegraphics[width=0.55\textwidth]{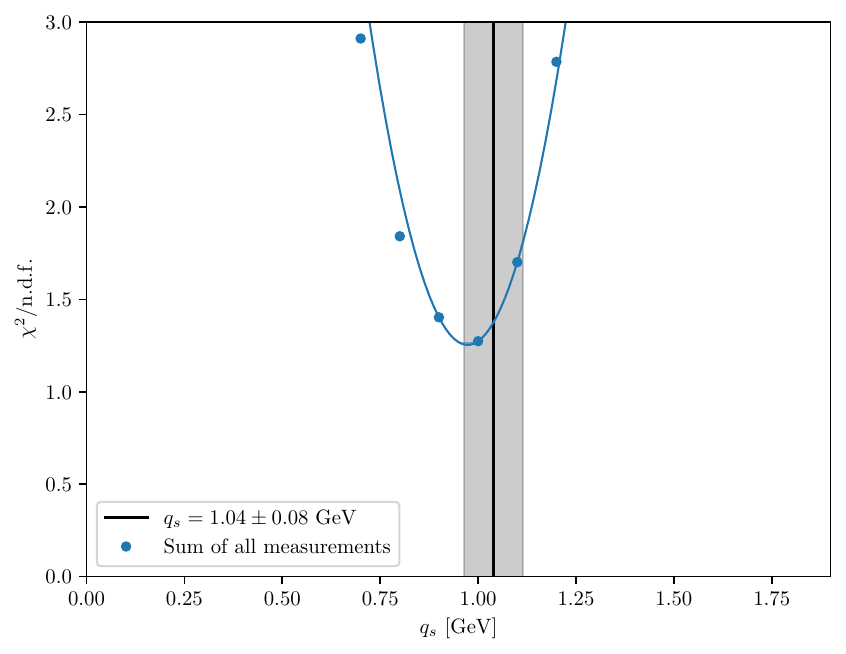}
 \caption{\small The reduced $\chi^2/\text{n.d.f.}$ distribution $(\text{n.d.f.} = 81$) as a function of $q_s$ obtained from a comparison of 
  the \CAS\ \PBset~Set2 prediction with the measurement of 
 Refs.~\protect\cite{CMS:2022ubq,LHCb:2021huf,CMS:2021ynu,Aad:2015auj,CDF:1999bpw,D0:1999jba,CDF:2012brb,Aidala:2018ajl,Moreno:1990sf}. 
 The minimum of global DY data fit is close to $q_s = 1$~\GeV\ and consistent with the CMS measurement~\protect\cite{CMS:2022ubq} shown separately by a black line. 
 }
\label{chi2ScanAll}
\end{center}
\end{figure}

The $\chi^2/\text{n.d.f}$ distribution as a function of $q_s$, for all the data together, is shown in Fig.~\ref{chi2ScanAll}. The $\chi^2$ distribution exhibits a minimum at around $q_s = 1.0 $ GeV, which is consistent with the value obtained as described above.

\begin{figure}[h!tb]
\begin{center} 
\includegraphics[width=0.49\textwidth]{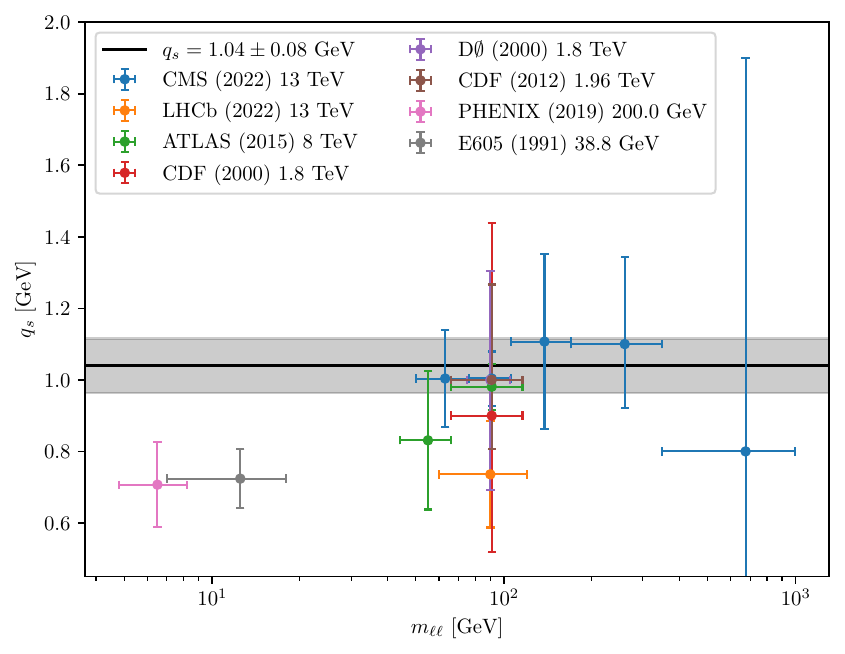}
\includegraphics[width=0.49\textwidth]{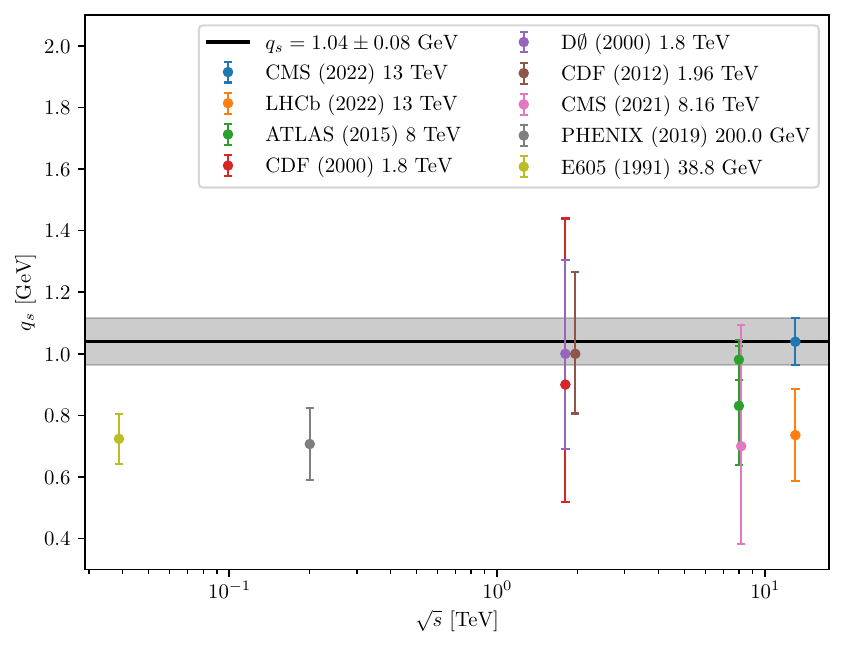}
 \caption{\small  
{\bf Left}: the value of $q_s$ as a function of the DY-mass as obtained from the measurements in 
  Refs.~\protect\cite{CMS:2022ubq,LHCb:2021huf,CMS:2021ynu,Aad:2015auj,CDF:1999bpw,D0:1999jba,CDF:2012brb,Aidala:2018ajl,Moreno:1990sf}. 
  {\bf Right} : same as a function of $\sqrt{s}$.
  The value of $q_s = 1.04 \pm 0.08$~\GeV\ as obtained from the measurements in Ref.~\protect\cite{CMS:2022ubq} is indicated.
 }
\label{qsversusmass}
\end{center}
\end{figure}

Figure~\ref{qsversusmass} displays the value of $q_s$ as a function of \mdy\ and $\sqrt{s}$ obtained from the different measurements in 
Refs.~\protect\cite{CMS:2022ubq,LHCb:2021huf,CMS:2021ynu,Aad:2015auj,CDF:1999bpw,D0:1999jba,CDF:2012brb,Aidala:2018ajl,Moreno:1990sf}.  
For the data which do not provide detailed uncertainty breakdown and are mainly used for the cross checks and comparison 
purpose, the uncertainty bars of $q_s$ shown in the figures are obtained from the $\chi^2$ variation of one unit and step
size of the $q_s$ scan only.
The value of $q_s=1.04 \pm 0.08$~\GeV\ , as derived from the measurements in Ref.\cite{CMS:2022ubq}, is
compatible for all 
ranges of \mdy, and also holds true for various values of $\sqrt{s}$. 
The obtained value is also found to be compatible for pPb data.

To summarize, we have obtained a value for the width of the Gauss distribution for modeling the intrinsic-$k_T$ distribution 
inside protons of $q_s = 1.04 \pm 0.08$ \GeV . This value, in contrast to standard Monte Carlo event generators, has no 
strong dependence on the center-of-mass energy as well as on 
the mass of the produced Drell-Yan lepton pair \mdy . 
The results of this section indicate that the treatment of soft emissions in the region $\zdyn  \lsim z < z_M$ with the strong coupling of Eq.~(\ref{freeze}) applied in \PBset~Set2 leads to intrinsic-\kt\ distributions with width parameter $q_s$ consistent with Fermi motion kinematics, and mildly varying with energy.

\section{Conclusion} \label{sec:concl} 

 In this paper we have carried out a detailed application of the \PBM-TMD
 methodology, which is reviewed in the first part of the paper, and used it 
to describe the DY low transverse momentum distributions across a wide range of DY masses.  
Within this methodology, we have presented the extraction of the  intrinsic-$k_T$  nonperturbative TMD parameter 
from fits to the measurements of DY  \pt\  differential cross sections performed recently 
at the LHC at $\sqrt{s}= 13$ TeV, for DY masses between 50 GeV and 1 TeV. 
We have compared this  with extractions from other DY measurements at different center-of-mass energies and masses. 

 As shown previously, the measured DY cross section at low \pt\ favours a choice of the strong coupling
$\alpha_s$ scale to be taken as the transverse momentum of each parton emission, as in angular-ordered CMW parton cascades. 
This corresponds to the TMD parton distribution set \PBset~Set2.
In this paper we use \PBset~Set2 with ``pre-confinement'' scale $q_0$ of 1 GeV. The  strong coupling is evaluated 
at the emitted transverse momentum $q_T$ for emissions
with $q_T > q_0$, populating the phase space region $ z < z_{\rm{dyn}}$ 
(where $z_{\rm{dyn}} = 1 - q_0 / | {\bf q}^{\prime } |$, with $| {\bf q}^{\prime } |$ being the scale of the branching),
while it is evaluated at the semi-hard scale $q_0$ for emissions with $q_T \lsim q_0$.
 The contribution to the Sudakov evolution from the parton branching in the phase space region $ z < z_{\rm{dyn}}$ gives the
perturbative resummation of Sudakov logarithms, while the contribution from the parton branching in the
phase space region $ z_{\rm{dyn}}  \lsim z < z_M$ gives the  nonperturbative Sudakov form factor. 
 Therefore the \PBset~Set2 contains two sources of nonperturbative 
effects: i) the nonperturbative TMD distribution at low evolution scale $\mu_0$, and ii) the nonperturbative
Sudakov form factor, specified by the ``pre-confinement'' scale  
prescription to continue the 
branching evolution to the infrared region $ z_{\rm{dyn}}  \lsim z < z_M$.   
The former includes the intrinsic-$k_T$ width parameter $q_s$, corresponding to Fermi motion in the 
hadron beam, while the latter is characterized by the semi-hard scale parameter $q_0$.
At low \kt, the contribution of nonperturbative Sudakov form factor interplays with the contribution of the 
intrinsic transverse momentum. 

 The main result of the present work is the extraction, within the \PBset~Set2  framework,  
of the intrinsic-$k_T$ Gauss distribution with zero mean and width parameter $q_s= \sqrt{2} \sigma$
 from the measured \pt\ dependence of the DY cross sections obtained recently at the LHC 
at $\sqrt{s}= 13$ TeV~\cite{CMS:2022ubq}, for different DY masses \mdy ,
between 50 GeV and 1 TeV. These measurements provide a complete decomposition of the different systematic 
uncertainties and their covariance matrices. 
To compare to the data, we have used DY production 
at NLO obtained with the \MCatNLO\ event generator
 matched with the \PBM\ TMD distributions \PBset~Set2,
with a given parameter $q_s$ value.
We performed a scan over a large range of values $q_s$ on the transverse momentum spectrum below the peak,
i.e.\ the sensitive part to intrinsic-\kt , and considering separately each experimental source of uncertainty and their
correlations. The theory scale uncertainties have been considered to be fully correlated inside each \mdy bin and
uncorrelated between \mdy bins. 
We found the value $q_s = 1.04 \pm 0.08$ \GeV , consistent for the different \mdy . 
The obtained value is in agreement to the expected value from Fermi-motion in protons.
It has been cross checked that this value is compatible with $q_s$ values obtained from other DY measurements at different
center-of-mass energies $\sqrt{s}$ and for a variety of DY masses. 
The global picture shows no strong dependence of the intrinsic-\kt\ on the center-of-mass energy or on the DY
mass, which contrasts with tuned standard Monte Carlo event generators that need a strongly increasing intrinsic Gauss width 
with $\sqrt{s}$ and with \mdy .

 We suggest that the remarkably stable value of $q_s$ that we obtain in our study can be 
attributed to the contribution of the nonperturbative Sudakov form factor and the treatment 
of the $ z_{\rm{dyn}}  \lsim z < z_M$ region near the soft-gluon resolution boundary.

\vskip 1 cm 
\begin{tolerant}{8000}
\noindent 
{\bf Acknowledgments.} 
We are grateful  to M. Abdullah Al-Mashad, 
L.I.~Estevez~Banos, L.~Keersmaekers, A.M.~van~Kampen,
K.~Wichmann,
Q.~Wang,
H.~Yang from the CASCADE developer group for interesting and productive discussions.
We are grateful to Louie Corpe for developing a code
to calculate $\chi^2$ including correlations within the 
Rivet frame. 
We also thank Marius Ambrozas for many discussion on the treatment of systematic uncertainties.  
This article is part of a project that has received funding from the European Union's Horizon 2020 research and 
innovation programme under grant agreement STRONG 2020 - No 824093.
LF is supported by the F.R.S.-FNRS of Belgium. A. L. acknowledges funding by Research Foundation-Flanders (FWO) (application number: 1272421N).
LM acknowledges the support of the Deutsche Forschungsgemeinschaft (DFG, German Research Foundation) under Germany's Excellence Strategy - EXC 2121 "Quantum Universe" - 390833306.
\end{tolerant} 
\vskip 0.6cm 

\bibliographystyle{mybibstyle-new.bst}
\raggedright  

\begin{thebibliography}{10}%
\makeatletter
\providecommand{\hrefCMSnoop }[0]{\@secondoftwo}%
\makeatother
\providecommand{\doi}{\texttt{doi:}\begingroup \urlstyle{tt}\Url}

\bibitem{Drell:1970wh}
\hrefCMSnoop {}{S.~Drell and T.-M. Yan, ``{Massive Lepton Pair Production in
  Hadron-Hadron Collisions at High-Energies}'',} \textit{ Phys. Rev. Lett.}
  \textbf{ 25} (1970) 316.

\bibitem{Collins:1984kg}
\hrefCMSnoop {}{J.~C. Collins, D.~E. Soper, and G.~F. Sterman, ``{Transverse
  Momentum Distribution in Drell-Yan Pair and W and Z Boson production}'',}
  \textit{ Nucl. Phys. B} \textbf{ 250} (1985)
199.

\bibitem{Sjostrand:2014zea}
T.~Sj{\"o}strand\hrefCMSnoop {}{ {et~al.}, ``{An introduction to PYTHIA
  8.2}'',} \textit{ Comput. Phys. Commun.} \textbf{ 191} (2015) 159,
\href{http://www.arXiv.org/abs/1410.3012}{\texttt{arXiv:1410.3012}}.

\bibitem{Bellm:2015jjp}
\hrefCMSnoop {}{J.~Bellm {et~al.}, ``{Herwig 7.0/Herwig++ 3.0 release note}'',}
  \textit{ Eur. Phys. J. C} \textbf{ 76} (2016) 196,
\href{http://www.arXiv.org/abs/1512.01178}{\texttt{arXiv:1512.01178}}.

\bibitem{Bahr:2008pv}
M.~Bahr\hrefCMSnoop {}{ {et~al.}, ``{Herwig++: physics and manual}'',} \textit{
  Eur. Phys. J. C} \textbf{ 58} (2008) 639--707,
  \href{http://www.arXiv.org/abs/0803.0883}{\texttt{arXiv:0803.0883}}.

\bibitem{Gleisberg:2008ta}
T.~Gleisberg\hrefCMSnoop {}{ {et~al.}, ``{Event generation with SHERPA 1.1}'',}
  \textit{ JHEP} \textbf{ 0902} (2009) 007,
\href{http://www.arXiv.org/abs/0811.4622}{\texttt{arXiv:0811.4622}}.

\bibitem{Martinez:2019mwt}
\hrefCMSnoop {}{A.~Bermudez~Martinez {et~al.}, ``{Production of Z-bosons in the
  parton branching method}'',} \textit{ Phys. Rev. D} \textbf{ 100} (2019)
  074027,
\href{http://www.arXiv.org/abs/1906.00919}{\texttt{arXiv:1906.00919}}.

\bibitem{Martinez:2020fzs}
\hrefCMSnoop {}{A.~Bermudez~Martinez {et~al.}, ``{The transverse momentum
  spectrum of low mass Drell--Yan production at next-to-leading order in the
  parton branching method}'',} \textit{ Eur. Phys. J. C} \textbf{ 80} (2020)
  598, \href{http://www.arXiv.org/abs/2001.06488}{\texttt{arXiv:2001.06488}}.

\bibitem{Hautmann:2017fcj}
F.~Hautmann\hrefCMSnoop {}{ {et~al.}, ``{Collinear and TMD quark and gluon
  densities from Parton Branching solution of QCD evolution equations}'',}
  \textit{ JHEP} \textbf{ 01} (2018) 070,
\href{http://www.arXiv.org/abs/1708.03279}{\texttt{arXiv:1708.03279}}.

\bibitem{Hautmann:2017xtx}
F.~Hautmann\hrefCMSnoop {}{ {et~al.}, ``{Soft-gluon resolution scale in QCD
  evolution equations}'',} \textit{ Phys. Lett. B} \textbf{ 772} (2017) 446,
\href{http://www.arXiv.org/abs/1704.01757}{\texttt{arXiv:1704.01757}}.

\bibitem{BermudezMartinez:2021lxz}
\hrefCMSnoop {}{A.~Bermudez~Martinez, F.~Hautmann, and M.~L. Mangano, ``{TMD
  evolution and multi-jet merging}'',} \textit{ Phys. Lett. B} \textbf{ 822}
  (2021) 136700,
  \href{http://www.arXiv.org/abs/2107.01224}{\texttt{arXiv:2107.01224}}.

\bibitem{Angeles-Martinez:2015sea}
\hrefCMSnoop {}{R.~Angeles-Martinez {et~al.}, ``{Transverse Momentum Dependent
  (TMD) parton distribution functions: status and prospects}'',} \textit{ Acta
  Phys. Polon. B} \textbf{ 46} (2015), no.~12, 2501,
\href{http://www.arXiv.org/abs/1507.05267}{\texttt{arXiv:1507.05267}}.

\bibitem{Gieseke:2007ad}
\hrefCMSnoop {}{S.~Gieseke, M.~H. Seymour, and A.~Siodmok, ``{A Model of
  non-perturbative gluon emission in an initial state parton shower}'',}
  \textit{ JHEP} \textbf{ 06} (2008) 001,
\href{http://www.arXiv.org/abs/0712.1199}{\texttt{arXiv:0712.1199}}.

\bibitem{Sjostrand:2004pf}
\hrefCMSnoop {}{T.~Sj\"ostrand and P.~Skands, ``{Multiple interactions and the
  structure of beam remnants}'',} \textit{ JHEP} \textbf{ 03} (2004) 053,
\href{http://www.arXiv.org/abs/hep-ph/0402078}{\texttt{arXiv:hep-ph/0402078}}.

\bibitem{Isaacson:2023iui}
\hrefCMSnoop {}{J.~Isaacson, Y.~Fu, and C.~P. Yuan, ``{Improving ResBos for the
  precision needs of the LHC}'',}
  \href{http://www.arXiv.org/abs/2311.09916}{\texttt{arXiv:2311.09916}}.

\bibitem{Hautmann:2020cyp}
\hrefCMSnoop {}{F.~Hautmann, I.~Scimemi, and A.~Vladimirov, ``{Non-perturbative
  contributions to vector-boson transverse momentum spectra in hadronic
  collisions}'',} \textit{ Phys. Lett. B} \textbf{ 806} (2020) 135478,
  \href{http://www.arXiv.org/abs/2002.12810}{\texttt{arXiv:2002.12810}}.

\bibitem{CMS:2022ubq}
\hrefCMSnoop {}{{CMS} Collaboration, ``{Measurement of the mass dependence of
  the transverse momentum of lepton pairs in Drell-Yan production in
  proton-proton collisions at $\sqrt{s}$ = 13 TeV}'',} \textit{ Eur. Phys. J.
  C} \textbf{ 83} (2023), no.~7, 628,
  \href{http://www.arXiv.org/abs/2205.04897}{\texttt{arXiv:2205.04897}}.

\bibitem{Bacchetta:2022awv}
A.~Bacchetta\hrefCMSnoop {}{ {et~al.}, ``{Unpolarized Transverse Momentum
  Distributions from a global fit of Drell-Yan and Semi-Inclusive
  Deep-Inelastic Scattering data}'',}
  \href{http://www.arXiv.org/abs/2206.07598}{\texttt{arXiv:2206.07598}}.

\bibitem{Bury:2022czx}
M.~Bury\hrefCMSnoop {}{ {et~al.}, ``{PDF bias and flavor dependence in TMD
  distributions}'',} \textit{ JHEP} \textbf{ 10} (2022) 118,
  \href{http://www.arXiv.org/abs/2201.07114}{\texttt{arXiv:2201.07114}}.

\bibitem{Martinez:2018jxt}
A.~Bermudez~Martinez\hrefCMSnoop {}{ {et~al.}, ``{Collinear and TMD parton
  densities from fits to precision DIS measurements in the parton branching
  method}'',} \textit{ Phys. Rev. D} \textbf{ 99} (2019) 074008,
\href{http://www.arXiv.org/abs/1804.11152}{\texttt{arXiv:1804.11152}}.

\bibitem{Jung:2021mox}
\hrefCMSnoop {}{H.~Jung, S.~T. Monfared, and T.~Wening, ``{Determination of
  collinear and TMD photon densities using the Parton Branching method}'',}
  \textit{ Physics Letters B} \textbf{ 817} (2021) 136299,
  \href{http://www.arXiv.org/abs/2102.01494}{\texttt{arXiv:2102.01494}}.

\bibitem{Jung:2021vym}
\hrefCMSnoop {}{H.~Jung and S.~T. Monfared, ``{TMD parton densities and
  corresponding parton showers: the advantage of four- and five-flavour
  schemes}'',}
  \href{http://www.arXiv.org/abs/2106.09791}{\texttt{arXiv:2106.09791}}.

\bibitem{BermudezMartinez:2020tys}
\hrefCMSnoop {}{A.~Bermudez~Martinez {et~al.}, ``{The transverse momentum
  spectrum of low mass Drell\textendash{}Yan production at next-to-leading
  order in the parton branching method}'',} \textit{ Eur. Phys. J. C} \textbf{
  80} (2020) 598,
  \href{http://www.arXiv.org/abs/2001.06488}{\texttt{arXiv:2001.06488}}.

\bibitem{Alwall:2014hca}
J.~Alwall\hrefCMSnoop {}{ {et~al.}, ``{The automated computation of tree-level
  and next-to-leading order differential cross sections, and their matching to
  parton shower simulations}'',} \textit{ JHEP} \textbf{ 1407} (2014) 079,
\href{http://www.arXiv.org/abs/1405.0301}{\texttt{arXiv:1405.0301}}.

\bibitem{Yang:2022qgk}
\hrefCMSnoop {}{H.~Yang {et~al.}, ``{Back-to-back azimuthal correlations in
  $\mathrm {Z} +$jet events at high transverse momentum in the TMD parton
  branching method at next-to-leading order}'',} \textit{ Eur. Phys. J. C}
  \textbf{ 82} (2022) 755,
  \href{http://www.arXiv.org/abs/2204.01528}{\texttt{arXiv:2204.01528}}.

\bibitem{Hautmann:2022xuc}
F.~Hautmann\hrefCMSnoop {}{ {et~al.}, ``{A parton branching with transverse
  momentum dependent splitting functions}'',} \textit{ Phys. Lett. B} \textbf{
  833} (2022) 137276,
  \href{http://www.arXiv.org/abs/2205.15873}{\texttt{arXiv:2205.15873}}.

\bibitem{Webber:1986mc}
\hrefCMSnoop {}{B.~R. Webber, ``{Monte Carlo Simulation of Hard Hadronic
  Processes}'',} \textit{ Ann. Rev. Nucl. Part. Sci.} \textbf{ 36} (1986)
253.

\bibitem{Marchesini:1987cf}
\hrefCMSnoop {}{G.~Marchesini and B.~R. Webber, ``{Monte Carlo Simulation of
  General Hard Processes with Coherent QCD Radiation}'',} \textit{ Nucl. Phys.
  B} \textbf{ 310} (1988)
461.

\bibitem{Catani:1990rr}
\hrefCMSnoop {}{S.~Catani, B.~R. Webber, and G.~Marchesini, ``{QCD coherent
  branching and semiinclusive processes at large x}'',} \textit{ Nucl. Phys. B}
  \textbf{ 349} (1991)
635.

\bibitem{Gribov:1972ri}
\hrefCMSnoop {}{V.~N. Gribov and L.~N. Lipatov, ``{Deep inelastic $e p$
  scattering in perturbation theory}'',} \textit{ Sov. J. Nucl. Phys.} \textbf{
  15} (1972) 438.
[Yad. Fiz.15,781(1972)].

\bibitem{Lipatov:1974qm}
\hrefCMSnoop {}{L.~N. Lipatov, ``{The parton model and perturbation theory}'',}
  \textit{ Sov. J. Nucl. Phys.} \textbf{ 20} (1975) 94.
[Yad. Fiz.20,181(1974)].

\bibitem{Altarelli:1977zs}
\hrefCMSnoop {}{G.~Altarelli and G.~Parisi, ``{Asymptotic freedom in parton
  language}'',} \textit{ Nucl. Phys. B} \textbf{ 126} (1977)
298.

\bibitem{Dokshitzer:1977sg}
\hrefCMSnoop {}{Y.~L. Dokshitzer, ``{Calculation of the structure functions for
  Deep Inelastic Scattering and $e^+ e^- $ annihilation by perturbation theory
  in Quantum Chromodynamics.}'',} \textit{ Sov. Phys. JETP} \textbf{ 46} (1977)
  641.
[Zh. Eksp. Teor. Fiz.73,1216(1977)].

\bibitem{Hautmann:2019biw}
\hrefCMSnoop {}{F.~Hautmann, L.~Keersmaekers, A.~Lelek, and A.~M. Van~Kampen,
  ``{Dynamical resolution scale in transverse momentum distributions at the
  LHC}'',} \textit{ Nucl. Phys. B} \textbf{ 949} (2019) 114795,
  \href{http://www.arXiv.org/abs/1908.08524}{\texttt{arXiv:1908.08524}}.

\bibitem{Abramowicz:2015mha}
\hrefCMSnoop {}{{ZEUS, H1} Collaboration, ``{Combination of measurements of
  inclusive deep inelastic ${e^{\pm }p}$ scattering cross sections and QCD
  analysis of HERA data}'',} \textit{ Eur. Phys. J. C} \textbf{ 75} (2015) 580,
\href{http://www.arXiv.org/abs/1506.06042}{\texttt{arXiv:1506.06042}}.

\bibitem{xFitterDevelopersTeam:2022koz}
{xFitter Developers' Team} Collaboration, \hrefCMSnoop {}{H.~Abdolmaleki
  {et~al.}, ``{xFitter: An Open Source QCD Analysis Framework. A resource and
  reference document for the Snowmass study}'',}
\newblock 6, 2022.
\newblock
  \href{http://www.arXiv.org/abs/2206.12465}{\texttt{arXiv:2206.12465}}.

\bibitem{Alekhin:2014irh}
\hrefCMSnoop {}{S.~Alekhin {et~al.}, ``\mbox{HERAFitter, Open Source QCD Fit
  Project}'',} \textit{ Eur. Phys. J. C} \textbf{ 75} (2015) 304,
\href{http://www.arXiv.org/abs/1410.4412}{\texttt{arXiv:1410.4412}}.

\bibitem{Hautmann:2014kza}
F.~Hautmann\hrefCMSnoop {}{ {et~al.}, ``{TMDlib and TMDplotter: library and
  plotting tools for transverse-momentum-dependent parton distributions}'',}
  \textit{ Eur. Phys. J. C} \textbf{ 74} (2014), no.~12, 3220,
\href{http://www.arXiv.org/abs/1408.3015}{\texttt{arXiv:1408.3015}}.

\bibitem{Abdulov:2021ivr}
\hrefCMSnoop {}{N.~A. Abdulov {et~al.}, ``{TMDlib2 and TMDplotter: a platform
  for 3D hadron structure studies}'',} \textit{ Eur. Phys. J. C} \textbf{ 81}
  (2021) 752,
  \href{http://www.arXiv.org/abs/2103.09741}{\texttt{arXiv:2103.09741}}.

\bibitem{Abdulhamid:2021xtt}
\hrefCMSnoop {}{M.~I. Abdulhamid {et~al.}, ``{Azimuthal correlations of high
  transverse momentum jets at next-to-leading order in the parton branching
  method}'',} \textit{ Eur. Phys. J. C} \textbf{ 82} (2022) 36,
  \href{http://www.arXiv.org/abs/2112.10465}{\texttt{arXiv:2112.10465}}.

\bibitem{Amati:1980ch}
D.~Amati\hrefCMSnoop {}{ {et~al.}, ``{A treatment of hard processes sensitive
  to the infrared structure of QCD}'',} \textit{ Nucl. Phys.} \textbf{ B173}
  (1980)
429.

\bibitem{Bassetto:1983mvz}
\hrefCMSnoop {}{A.~Bassetto, M.~Ciafaloni, and G.~Marchesini, ``{Jet Structure
  and Infrared Sensitive Quantities in Perturbative QCD}'',} \textit{ Phys.
  Rept.} \textbf{ 100} (1983) 201--272.

\bibitem{vanKampen:2021oxe}
\hrefCMSnoop {}{A.~M. van Kampen, ``{Drell-Yan transverse spectra at the LHC: a
  comparison of parton branching and analytical resummation approaches}'',}
  \textit{ SciPost Phys. Proc.} \textbf{ 8} (2022) 151,
  \href{http://www.arXiv.org/abs/2108.04099}{\texttt{arXiv:2108.04099}}.

\bibitem{PB-NNLL}
A.~\mbox{Bermudez~Martinez}\hrefCMSnoop {}{ {et~al.}} to be published.

\bibitem{PBevolution}
\hrefCMSnoop {}{{H. Jung et al.}, ``{The Parton Branching evolution for
  collinear and TMD parton densities - uPDFevolv2}''.} to be published, 2023.

\bibitem{Nagy:2020gjv}
\hrefCMSnoop {}{Z.~Nagy and D.~E. Soper, ``{Evolution of parton showers and
  parton distribution functions}'',} \textit{ Phys. Rev. D} \textbf{ 102}
  (2020), no.~1, 014025,
  \href{http://www.arXiv.org/abs/2002.04125}{\texttt{arXiv:2002.04125}}.

\bibitem{Frixione:2023ssx}
\hrefCMSnoop {}{S.~Frixione and B.~R. Webber, ``{Correcting for cutoff
  dependence in backward evolution of QCD parton showers}'',}
  \href{http://www.arXiv.org/abs/2309.15587}{\texttt{arXiv:2309.15587}}.

\bibitem{Mendizabal:2023mel}
\hrefCMSnoop {}{M.~Mendizabal, F.~Guzman, H.~Jung, and S.~Taheri~Monfared,
  ``{On the role of soft gluons in collinear parton densities}'',}
  \href{http://www.arXiv.org/abs/2309.11802}{\texttt{arXiv:2309.11802}}.

\bibitem{Baranov:2021uol}
\hrefCMSnoop {}{S.~Baranov {et~al.}, ``{CASCADE3 A Monte Carlo event generator
  based on TMDs}'',} \textit{ Eur. Phys. J. C} \textbf{ 81} (2021) 425,
  \href{http://www.arXiv.org/abs/2101.10221}{\texttt{arXiv:2101.10221}}.

\bibitem{Sjostrand:2006za}
\hrefCMSnoop {}{T.~Sj\"ostrand, S.~Mrenna, and P.~Skands, ``{PYTHIA 6.4 physics
  and manual}'',} \textit{ JHEP} \textbf{ 05} (2006) 026,
\href{http://www.arXiv.org/abs/hep-ph/0603175}{\texttt{arXiv:hep-ph/0603175}}.

\bibitem{Sirunyan:2018owv}
\hrefCMSnoop {}{{CMS} Collaboration, ``{Measurement of the differential
  Drell-Yan cross section in proton-proton collisions at $ \sqrt{\mathrm{s}} $
  = 13 TeV}'',} \textit{ JHEP} \textbf{ 12} (2019) 059,
  \href{http://www.arXiv.org/abs/1812.10529}{\texttt{arXiv:1812.10529}}.

\bibitem{Buckley:2010ar}
A.~Buckley\hrefCMSnoop {}{ {et~al.}, ``{Rivet user manual}'',} \textit{ Comput.
  Phys. Commun.} \textbf{ 184} (2013) 2803,
\href{http://www.arXiv.org/abs/1003.0694}{\texttt{arXiv:1003.0694}}.

\bibitem{Sirunyan:2019bzr}
\hrefCMSnoop {}{{CMS} Collaboration, ``{Measurements of differential Z boson
  production cross sections in proton-proton collisions at $ \sqrt{s} $ = 13
  TeV}'',} \textit{ JHEP} \textbf{ 12} (2019) 061,
\href{http://www.arXiv.org/abs/1909.04133}{\texttt{arXiv:1909.04133}}.

\bibitem{Aad:2014qja}
\hrefCMSnoop {}{{ATLAS} Collaboration, ``{Measurement of the low-mass Drell-Yan
  differential cross section at $\sqrt{s}$ = 7 TeV using the ATLAS
  detector}'',} \textit{ JHEP} \textbf{ 06} (2014) 112,
\href{http://www.arXiv.org/abs/1404.1212}{\texttt{arXiv:1404.1212}}.

\bibitem{CMS:2015vap}
\hrefCMSnoop {}{{CMS} Collaboration, ``{Study of Final-State Radiation in
  Decays of Z Bosons Produced in $pp$ Collisions at 7 TeV}'',} \textit{ Phys.
  Rev. D} \textbf{ 91} (2015) 092012,
  \href{http://www.arXiv.org/abs/1502.07940}{\texttt{arXiv:1502.07940}}.

\bibitem{BermudezMartinez:2022bpj}
\hrefCMSnoop {}{A.~Bermudez~Martinez, F.~Hautmann, and M.~L. Mangano,
  ``{Multi-jet merging with TMD parton branching}'',} \textit{ JHEP} \textbf{
  09} (2022) 060,
  \href{http://www.arXiv.org/abs/2208.02276}{\texttt{arXiv:2208.02276}}.

\bibitem{Martinez:2021dwx}
\hrefCMSnoop {}{A.~Bermudez~Martinez, F.~Hautmann, and M.~L. Mangano,
  ``{Multi-jet physics at high-energy colliders and TMD parton evolution}'',}
\newblock 2021.
\newblock
  \href{http://www.arXiv.org/abs/2109.08173}{\texttt{arXiv:2109.08173}}.

\bibitem{LHCb:2021huf}
\hrefCMSnoop {}{{LHCb} Collaboration, ``{Precision measurement of forward $Z$
  boson production in proton-proton collisions at $\sqrt{s} = 13$ TeV}'',}
  \textit{ JHEP} \textbf{ 07} (2022) 026,
  \href{http://www.arXiv.org/abs/2112.07458}{\texttt{arXiv:2112.07458}}.

\bibitem{Chi2LMoureaux}
\href {https://github.com/lmoureaux/CovarianceFits}{L.~Moureaux and I.~Bubanja,
  ``Fits with covariance matrices''.}
  https://github.com/lmoureaux/CovarianceFits.

\bibitem{lhcew-LCorpe}
\href
  {https://gitlab.cern.ch/lhcewkwg/lhcewkwg-vjets/correlations-library}{L.~Corpe,
  ``Correlations Library''.} Contribution to yellow report of LHCEW working
  group: Jet and electroweak bosons, 2019.

\bibitem{Aad:2015auj}
\hrefCMSnoop {}{{ATLAS} Collaboration, ``{Measurement of the transverse
  momentum and $\phi ^*_{\eta }$ distributions of Drell--Yan lepton pairs in
  proton--proton collisions at $\sqrt{s}=8$ TeV with the ATLAS detector}'',}
  \textit{ Eur. Phys. J. C} \textbf{ 76} (2016) 291,
\href{http://www.arXiv.org/abs/1512.02192}{\texttt{arXiv:1512.02192}}.

\bibitem{D0:1999jba}
\hrefCMSnoop {}{{D0} Collaboration, ``{Measurement of the inclusive
  differential cross section for $Z$ bosons as a function of transverse
  momentum in $\bar{p}p$ collisions at $\sqrt{s} = 1.8$ TeV}'',} \textit{ Phys.
  Rev. D} \textbf{ 61} (2000) 032004,
  \href{http://www.arXiv.org/abs/hep-ex/9907009}{\texttt{arXiv:hep-ex/9907009}}.

\bibitem{CDF:1999bpw}
\hrefCMSnoop {}{{CDF} Collaboration, ``{The transverse momentum and total cross
  section of $e^+e^-$ pairs in the $Z$ boson region from $p\bar{p}$ collisions
  at $\sqrt{s} = 1.8$ TeV}'',} \textit{ Phys. Rev. Lett.} \textbf{ 84} (2000)
  845,
  \href{http://www.arXiv.org/abs/hep-ex/0001021}{\texttt{arXiv:hep-ex/0001021}}.

\bibitem{CDF:2012brb}
\hrefCMSnoop {}{{CDF} Collaboration, ``{Transverse momentum cross section of
  $e^+e^-$ pairs in the $Z$-boson region from $p\bar{p}$ collisions at
  $\sqrt{s}=1.96$ TeV}'',} \textit{ Phys. Rev. D} \textbf{ 86} (2012) 052010,
  \href{http://www.arXiv.org/abs/1207.7138}{\texttt{arXiv:1207.7138}}.

\bibitem{Aidala:2018ajl}
\hrefCMSnoop {}{{PHENIX} Collaboration, ``{Measurements of $\mu\mu$ pairs from
  open heavy flavor and Drell-Yan in $p+p$ collisions at $\sqrt{s}=200$
  GeV}'',} \textit{ Phys. Rev. D} \textbf{ 99} (2019) 072003,
\href{http://www.arXiv.org/abs/1805.02448}{\texttt{arXiv:1805.02448}}.

\bibitem{Moreno:1990sf}
\hrefCMSnoop {}{G.~Moreno {et~al.}, ``{Dimuon production in proton - copper
  collisions at $\sqrt{s}$ = 38.8~GeV}'',} \textit{ Phys. Rev. D} \textbf{ 43}
  (1991)
2815.

\bibitem{CMS:2021ynu}
\hrefCMSnoop {}{{CMS} Collaboration, ``{Study of Drell-Yan dimuon production in
  proton-lead collisions at $\sqrt{s_\mathrm{NN}} =$ 8.16 TeV}'',} \textit{
  JHEP} \textbf{ 05} (2021) 182,
  \href{http://www.arXiv.org/abs/2102.13648}{\texttt{arXiv:2102.13648}}.

\end{thebibliography}
\providecommand{\href}[2]{#2}\begingroup\raggedright\endgroup

\end{document}